\magnification\magstep 1
\parskip 4pt plus 1pt minus 0.5pt
\def\init{\tabskip 0pt}
\def\crr{\cr\noalign{\hrule}}
\def\abs#1{\vert{#1}\vert}
\def\arg{\mathop{\rm Arg}}
\def\d{{\rm d}}
\def\eps{\varepsilon}
\def\frac#1#2{{#1\over#2}}
\def\im{\mathop{\rm Im}}
\def\l{\lambda}
\def\min{{\rm min}}
\def\max{{\rm max}}
\def\r#1{{\rm#1}}
\def\re{\mathop{\rm Re}}
\def\w{\widetilde}
\def\x{{\bf x}}
\def\y{{\bf y}}
\def\A{{\bf A}}
\def\B{{\bf B}}
\def\Disc{\mathop{\rm Disc}}
\def\EMA{{\rm EMA}}
\def\LN{_{\rm LN}}
\def\L{^{(L)}}
\def\M{{\bf M}}
\def\P{{\cal P}}
\def\Q{{\cal Q}}
\def\R{^{(R)}}
\def\V{{\bf V}}
\def\X{{\bf X}}
\def\0{{\bf 0}}
%
\centerline{DIELECTRIC RESONANCES OF BINARY RANDOM NETWORKS}
\vskip 30pt
\rm
\centerline{by Th. Jonckheere$^{(1,\r{a})}$ and J.M. Luck$^{(2,\r{b})}$}
\vskip 30pt
\noindent (1) Laboratoire Kastler-Brossel, 4, Place Jussieu,
Tour 12, 1er~\'etage, 75252 Paris cedex~05, France.
\smallskip
\noindent (2) C.E.A. Saclay, Service de Physique Th\'eorique,
91191 Gif-sur-Yvette cedex, France.
\vskip 50pt
\noindent{\bf Abstract.}
We investigate the AC conductivity of binary random impedance networks,
with emphasis on its dependence on the ratio $h=\sigma_1/\sigma_0$,
with $\sigma_0$ and $\sigma_1$ being the complex conductances of both phases,
occurring with respective probabilities $p$ and $1-p$.
We propose an algorithm to determine the rational
$h$-dependence of the conductance of a finite network,
in terms of its poles and of the associated residues.
The poles, which lie on the negative real $h$-axis,
are called resonances, since they show up as narrow resonances
in the AC conductance of the $RL-C$ model of a metal-dielectric composite
with a high quality factor $Q$.
This approach is an extension of a previous work
devoted to the dielectric resonances of isolated finite clusters.
A numerical implementation of the algorithm,
on the example of the square lattice,
allows a detailed investigation of the resonant dielectric
response of the binary model, including the $p$-dependence
of the density of resonances and the associated spectral function,
the Lifshitz behaviour of these quantities near the endpoints of the spectrum
of resonances, the distribution of spacings between neighbouring resonances,
and the $Q$-dependence of the fraction of visible resonances
in the $RL-C$ model.
The distribution of the local electric fields at resonance
is found to be multifractal.
This result is put in perspective with the giant
surface-enhanced Raman scattering observed e.g. in semicontinuous metal films.
\vfill
\noindent(a) e-mail: thib@spectro.jussieu.fr

\noindent(b) Corresponding author. e-mail: luck@spht.saclay.cea.fr
\eject
\noindent {\bf 1 INTRODUCTION}
\smallskip

Random networks of complex impedances are currently used to model
electrical and optical properties of disordered inhomogeneous media.
The most common situation is that of a binary composite medium,
modeled by attributing a random conductance to each bond $(\x,\y)$
of a lattice, according to the binary law
$$
\sigma_{\x,\y}=\left\{\matrix{
\sigma_0&\hbox{with probability}&p,\hfill\cr
\sigma_1&\hbox{with probability}&q=1-p,\hfill\cr
}\right.
\eqno(1.1)
$$
in correspondence with the bond percolation problem
(see ref.~[1] for a review).

The conductances (inverse impedances, or admittances)
$\sigma_0$ and $\sigma_1$ of both phases take arbitrary
frequency-dependent complex values.
Hereafter we follow the notations of ref.~[2].
The dimensionless complex ratio
$$
h=\frac{\sigma_1}{\sigma_0}
\eqno(1.2)
$$
and the concentration $p$ are the essential parameters of the model.
As far as static (DC) properties are concerned, the limiting case $h=0$
embraces the conductor-insulator mixture $(\sigma_1=0)$
and the superconductor-conductor mixture $(\sigma_0=\infty)$.
In both situations the conductivity exhibits power-law behaviour
for $p$ close to the critical concentration $p_c$,
corresponding to the percolation threshold.

More generally, the conductivity of the binary model
obeys a scaling law in the critical regions
defined by $\abs{h}\ll 1$ and $\abs{p-p_c}\ll 1$,
or equivalently $\abs{h}\gg 1$ and $\abs{q-p_c}\ll 1$~[1].
Frequency-dependent (AC) properties of metal-dielectric composites,
such as cermets or thin films,
are often investigated by means of either the $R-C$ and the $RL-C$ models~[3].
In both cases the low-frequency regime corresponds to $\abs{h}\gg 1$,
and thus to critical behaviour when the metallic concentration $q$
is near the percolation threshold.

The purpose of the present work
is to shed some new light on the resonant behaviour of the binary model.
The emphasis will be put on the analytic structure
of the conductivity in the ratio $h$, or in the equivalent complex variable
$$
\l=\frac{1}{1-h}=\frac{\sigma_0}{\sigma_0-\sigma_1}.
\eqno(1.3)
$$
For any value of the concentration $p$,
the conductivity has singularities for $h$ real and negative,
i.e., in the range $0\le\l\le 1$.
A quantitative investigation of these singularities,
and of related quantities, is the main goal of the present paper.
Our aim is twofold.
First, the analytic structure of the conductivity
of the binary model is a classical subject,
since the developments of the Bergman-Milton theory~[4, 5].
The key ingredient of this formalism is the spectral function $H(p,x)$,
which has only been the subject of a limited number of investigations
so far~[6, 7].
The present approach provides a direct accurate numerical evaluation of the
spectral function of binary random networks.
Second, the singularities of the conductivity have a physical interpretation
in terms of dielectric resonances in the $RL-C$ model,
and of the relaxation times in the transient response in the $R-C$ model.
The regime of dielectric resonances has been argued to provide a
natural explanation
for the anomalous fluctuations of the local electric field~[8],
which are responsible for giant surface-enhanced Raman scattering
observed e.g. in semicontinuous metal films~[9].

The dielectric resonances of isolated clusters
have been investigated in ref.~[2].
The situation considered there was a finite set
of (metallic) bonds with conductance $\sigma_1$,
embedded in an infinite (dielectric) host lattice,
whose bonds have a conductance $\sigma_0$.
It was shown there that the conductance of such a system
is entirely characterised by a finite number of resonances.
The positions $\l_a$ of the resonances,
and the associated cross-sections $\gamma_a$,
are expressed in terms of the eigenvalues and eigenvectors
of a finite matrix $\M$, dictated by the geometry of the clusters.
The present paper is an extension of
the method of ref.~[2] to an arbitrary binary network.
We shall make use of an efficient algorithm,
which allows for an exact determination of all the resonances
of a finite sample.
The setup of this paper is as follows.
In section~2, we gather definitions and results
on various features of the conductivity of the binary model.
Section~3 is devoted to the presentation of the algorithm.
Section~4 contains a variety of numerical results, concerning especially
the spectral function of the Bergman-Milton theory
and the spectral density of resonances,
their Lifshitz behaviour near the endpoints of the spectrum
($\l=0$ and $\l=1$), the number of visible resonances of a finite sample
and its dependence on the quality factor $Q$ of the $RL-C$ model,
and the distribution of the local electric fields at resonance,
which turns out to be multifractal.

\medskip
\noindent {\bf 2 THEORETICAL BACKGROUND}
\smallskip
\noindent {\bf 2.1 Conductance and impedance of a finite network}
\smallskip

Consider the binary model
on a finite network of size $M\times N$, as shown in Figure~1.
This network contains $n_S=(M-1)N$ sites (nodes)
and $n_B=MN+(M-1)(N-1)$ bonds (links),
among which $MN$ are horizontal (perpendicular to the electrodes),
and $(M-1)(N-1)$ are vertical (parallel to the electrodes).
The bonds with a conductance $\sigma_0$ are called the $\P$-bonds,
which form the $\P$-set.
There are $n_\P$ of them,
among which $n_\P^H$ are horizontal and $n_\P^V$ are vertical.
Similarly, the bonds with a conductance $\sigma_1$ are called the $\Q$-bonds,
which form the $\Q$-set.
There are $n_\Q$ of them,
among which $n_\Q^H$ are horizontal and $n_\Q^V$ are vertical.

Let $Y$ be the conductance (admittance) of the network,
measured between the plane electrodes (bus bars) shown in Figure~1,
and let $Z=1/Y$ be its impedance.
These quantities
are rational functions of the dimensionless complex variables $h$ or $\l$.
Let us anticipate that it is more convenient to use the variable $\l$.
We have
$$
\eqalignno{
Y&=\frac{N\sigma_0}{M}\prod_{a=1}^{n_R}\frac{\l-\w\l_a}{\l-\l_a}
=\frac{N\sigma_0}{M}\left(1-\sum_{a=1}^{n_R}\frac{\alpha_a}{\l-\l_a}\right),
&(2.1\r{a})\cr
Z&=\frac{M}{N\sigma_0}\prod_{a=1}^{n_R}\frac{\l-\l_a}{\l-\w\l_a}
=\frac{M}{N\sigma_0}\left(1+\sum_{a=1}^{n_R}\frac{\beta_a}{\l-\w\l_a}\right).
&(2.1\r{b})\cr
}
$$
The prefactors of these expressions
are the conductance and the impedance of the uniform network
whose bonds have a conductance $\sigma_0$,
since $\l=\infty$ corresponds to $\sigma_1=\sigma_0$ (no disorder).
The $\l_a$ involved in the product expressions
are the poles of the conductance and the zeros of the impedance,
while the $\w\l_a$ are the zeros of the conductance
and the poles of the impedance.
The number $n_R$ of poles and zeros depends on the configuration
of the random bonds.
It is bounded by the number of sites of the network: $0\le n_R\le n_S$.
Finally, the partial-fraction expansions
involve residues $\alpha_a$ and $\beta_a$.

It follows from considerations on the dissipated power
that $Y/\sigma_0$ is a Stieltjes function,
namely its imaginary part has the sign of $\im h$,
or equivalently of $\im\l$~[4, 5, 1].
This property implies that the poles and zeros of $Y$ and of $Z$ alternate,
according to
$$
0\le\l_1<\w\l_1<\cdots<\l_{n_R}<\w\l_{n_R}\le 1.
\eqno(2.2)
$$
We have $\l_1=0$ if, and only if, the $\Q$-set is conducting,
i.e., it contains at least one connected path between both electrodes.
The conductance of this $\Q$-set then reads
$$
Y_\Q=\frac{N}{M}\alpha_1\sigma_1.
\eqno(2.3)
$$
Similarly, we have $\w\l_{n_R}=1$ if, and only if,
the $\P$-set is not conducting.

The Stieltjes property also implies that the residues
$\alpha_a$ and $\beta_a$ are positive.
These two sets of residues are different from each other in general.
They obey, however, a remarkable sum rule,
which can be proved by expanding the conductance $Y$ around $\l=\infty$,
to first order in $1/\l$.
Indeed, $1/\l=(\sigma_0-\sigma_1)/\sigma_0$
is the dimensionless contrast between the conductances of both phases.
As a consequence, for a given realisation of the random network,
up to first order in $1/\l$ included,
the conductance reads $Y=(N/M)\overline{\sigma_H}$, where
$$
\overline{\sigma_H}=\frac{n_\P^H\sigma_0+n_\Q^H\sigma_1}{MN}
=\sigma_0\left(1-\frac{n_\Q^H}{MN\,\l}\right)
\eqno(2.4)
$$
is the average value of the conductances of the horizontal bonds
of the network.
By inserting this estimate into the second expressions of eqs.~(2.1a, 1b),
we obtain the relation
$$
\sum_{a=1}^{n_R}\alpha_a=\sum_{a=1}^{n_R}\beta_a=\frac{n_\Q^H}{MN}.
\eqno(2.5)
$$

In the case of isolated finite clusters~[2],
the poles and zeros of the conductance form very tight doublets.
We have indeed
$$
\alpha_a\approx\beta_a\approx\w\l_a-\l_a\approx\frac{\gamma_a\l_a(1-\l_a)}{MN}.
\eqno(2.6)
$$
The sum rule~(2.5) then becomes
$$
\sum_{a=1}^{n_R}\gamma_a\l_a(1-\l_a)=n_H,
\eqno(2.7)
$$
where $n_H$ is the number of horizontal bonds in the clusters
(perpendicular to the electrodes).
This identity was not noticed in ref.~[2].

\medskip
\noindent {\bf 2.2 $R-C$ model}
\smallskip

The $R-C$ model, already mentioned in the Introduction, is defined as follows.
The $\Q$-set is a metallic phase, whose bonds consist of a pure resistance $R$,
while the $\P$-set is a dielectric phase,
whose bonds consist of a perfect capacitance $C$.
The complex conductances at frequency $f=\omega/(2\pi)$ thus read
$$
\sigma_0=iC\omega,\qquad\sigma_1=\frac{1}{R}.
\eqno(2.8)
$$
Along the lines of refs.~[1, 2], we introduce the microscopic relaxation time
$$
\tau=RC,
\eqno(2.9)
$$
so that
$$
h=\frac{1}{i\omega\tau},\qquad\l=\frac{i\omega\tau}{i\omega\tau-1}.
\eqno(2.10)
$$
We thus have $h\to\infty$ and $\l\to 0$ at low frequency.

The poles of the conductance $Y$
show up as relaxation times in the transient response of the model~[1].
Consider indeed the $R-C$ model on a finite network,
submitted to a voltage $V(t)=V_0\theta(t)$ applied between the electrodes,
with $\theta(t)$ being the Heaviside step-function.
The intensity $I(t)$ across the network in the transient regime
can be evaluated by means of the Fourier transformation.
We thus obtain
$$
I(t)=\int_{-\infty}^{+\infty}\frac{\d\omega}{2\pi}Y(\omega)e^{i\omega t}
\frac{V_0}{i\omega+0}.
\eqno(2.11)
$$
The second expression of eq.~(2.1a) for the conductance $Y$,
with $\l$ given in eq.~(2.10), yields the result
$$
I(t)=\frac{NV_0}{MR}\sum_{a=1}^{n_R}\frac{\alpha_a}{(1-\l_a)^2}
\exp\left(-\frac{t}{\tau_a}\right),
\eqno(2.12)
$$
where the relaxation times read
$$
\tau_a=\frac{1-\l_a}{\l_a}\tau.
\eqno(2.13)
$$
If the metallic $\Q$-phase is conducting, we have $\l_1=0$.
The term $a=1$ in eq.~(2.12) yields the DC current through the network,
$I_0=NV_0\alpha_1/(MR)$, in agreement with the result~(2.3).

\medskip
\noindent {\bf 2.3 $RL-C$ model}
\smallskip

In the $RL-C$ model, already mentioned in the Introduction,
the metallic bonds of the $\Q$-set
now consist of an inductance $L$ in series with a weak resistance $R$,
while the dielectric bonds of the $\P$-set
still consist of a perfect capacitance $C$.
The conductances at frequency $f=\omega/(2\pi)$ now read
$$
\sigma_0=iC\omega,\qquad\sigma_1=\frac{1}{R+iL\omega}.
\eqno(2.14)
$$
Along the lines of refs.~[1--3], we introduce the microscopic
resonance frequency
$$
\omega_0=\frac{1}{\sqrt{LC}},
\eqno(2.15)
$$
the reduced frequency
$$
y=\frac{\omega}{\omega_0},
\eqno(2.16)
$$
and the quality factor
$$
Q=\frac{1}{R}\sqrt{\frac{L}{C}}=\frac{L\omega_0}{R}=\frac{1}{RC\omega_0},
\eqno(2.17)
$$
which is a dimensionless measure of the dissipation rate.

In the following, we shall mostly consider the case of a weak dissipation,
corresponding to a large quality factor $(Q\gg 1)$.
We have then
$$
h=\frac{1}{-y^2+iy/Q}\approx -\frac{1}{y^2}-\frac{i}{y^3Q},\qquad
\l=\frac{y^2-iy/Q}{1+y^2-iy/Q}\approx\frac{y^2}{1+y^2}-\frac{iy}{(1+y^2)^2Q}.
\eqno(2.18)
$$
We notice that the low-frequency regime $(y\to0)$
again corresponds to $h\to\infty$ and $\l\to 0$.

Since the variables $h$ and $\l$ have a small (negative) imaginary part,
proportional to $1/Q$, the poles of the conductance $Y$
and of the impedance $Z$ of the network show up
in the frequency dependence of these quantities as narrow resonances,
respectively located at $\omega_a=\omega_0y_a$
and $\w\omega_a=\omega_0\w y_a$, with
$$
y_a=\sqrt{\frac{\l_a}{1-\l_a}},\qquad\w y_a=\sqrt{\frac{\w\l_a}{1-\w\l_a}}.
\eqno(2.19)
$$
We have indeed
$$
\eqalignno{
Y&\approx\frac{NC\omega_0}{2M}\sum_{a=1}^{n_R}
\frac{\alpha_a}{(1-\l_a)^2}\,\frac{1/(2Q)-i(y-y_a)}{(y-y_a)^2+1/(4Q^2)},
&(2.20\r{a})\cr
Z&\approx\frac{M}{2NC\omega_0}\sum_{a=1}^{n_R}
\frac{\beta_a}{\w\l_a(1-\w\l_a)}
\,\frac{1/(2Q)-i(y-\w y_a)}{(y-\w y_a)^2+1/(4Q^2)}.
&(2.20\r{b})\cr
}
$$

Eq.~(2.20a) shows that the real part $\re Y(\omega)$
of the conductance exhibits $n_R$ narrow resonances, at $\omega_a=\omega_0y_a$.
The resonance peaks have a Lorentzian shape,
with a common absolute width $\Delta\omega=\omega_0/(2Q)$.
The maxima at resonance read
$$
\big(\re Y\big)_\max\approx\frac{N}{MR}\,\frac{\alpha_a}{(1-\l_a)^2},
\eqno(2.21)
$$
while the area under each resonance peak is
$$
{\cal A}_a=\int_{\omega\sim\omega_a}\re Y(\omega)\d\omega
\approx\frac{\pi N}{2ML}\,\frac{\alpha_a}{(1-\l_a)^2}.
\eqno(2.22)
$$
A similar pattern of resonances
can be observed on the real part $\re Z(\omega)$ of the im\-pe\-dan\-ce~(2.20b),
with maxima at resonance and areas under resonance peaks respectively given by
$$
\big(\re Z\big)_\max
\approx\frac{ML}{NRC}\,\frac{\beta_a}{\w\l_a(1-\w\l_a)},\qquad
\w{\cal A}_a\approx\frac{\pi M}{2NC}\,\frac{\beta_a}{\w\l_a(1-\w\l_a)}.
\eqno(2.23)
$$

\medskip
\noindent {\bf 2.4 Spectral function and density of resonances}
\smallskip

The conductance $Y$ of an infinitely large network
drawn on any regular lattice becomes a self-averaging quantity.
In other words, the conductivity
$$
\Sigma(p,\l)=\lim_{M,N\to\infty}\frac{M\,Y}{N}
\eqno(2.24)
$$
is an intrinsic characteristic of the binary model,
which depends on the lattice under consideration,
on the concentration $p$, and on the complex variable $h$ or $\l$.
The conductivity of the binary model has been the subject
of many investigations~[1].
We gather below some definitions and properties
which will be useful in the sequel.

The second expression of eq.~(2.1a) yields the Bergman-Milton
integral representation~[4, 5]
$$
\Sigma(p,\l)=\sigma_0\left(1-\int_0^1\frac{H(p,x)\d x}{\l-x}\right),
\eqno(2.25)
$$
where the positive function
$$
H(p,x)=\lim_{M,N\to\infty}\sum_{a=1}^{n_R}\alpha_a\delta(x-\l_a)
\eqno(2.26)
$$
is called the spectral function of the binary model.
In more precise terms, it is the density of a positive measure,
supported by the interval $0\le x\le 1$.
Indeed, as a consequence of eq.~(2.3),
the spectral function has a singular component at $x=0$, of the form
$$
H_\r{sg}(p,x)=A(1-p)\,\delta(x),
\eqno(2.27)
$$
with the notation to be introduced in eq.~(2.38),
whenever the $\Q$-set is conducting, i.e., for $1-p=q>p_c$.

The spectral function entirely determines the conductivity $\Sigma(p,\l)$,
by means of the integral representation~(2.25).
Conversely, it is given by the inverse formula
$$
H(p,x)=\frac{1}{\pi}\im\frac{\Sigma(p,x+i0)}{\sigma_0}.
\eqno(2.28)
$$
In other words, the conductivity is analytic in the complex $\l$-plane
cut along the interval $0\le\l\le 1$,
and its discontinuity along this cut reads
$$
\Disc\Sigma(p,\l)=2i\pi\sigma_0H(p,\l).
\eqno(2.29)
$$

The spectral function directly yields the transient intensity
$I(t)$ of the $R-C$ model on a very large network, namely
$$
I(t)\approx\frac{NV_0}{MR}\int_0^1\frac{H(p,x)\d x}{(1-x)^2}
\exp\left(-\frac{x}{1-x}\frac{t}{\tau}\right).
\eqno(2.30)
$$

To close up, we introduce the spectral density of resonances
$$
\rho(p,x)=\lim_{M,N\to\infty}\frac{1}{MN}\sum_{a=1}^{n_R}\delta(x-\l_a),
\eqno(2.31)
$$
and the total density of resonances
$$
\rho_R(p)=\lim_{M,N\to\infty}\frac{n_R}{MN}=\int_0^1\rho(p,x)\d x,
\eqno(2.32)
$$
representing the mean number of resonances per site.
This quantity has a non-trivial dependence on the concentration $p$.
It will be expressed in eq.~(3.14)
in terms of geometrical quantities of the bond percolation problem.

\medskip
\noindent {\bf 2.5 Homogeneity and duality}
\smallskip

First, the conductivity of the binary model
is invariant under the simultaneous interchange $p\leftrightarrow q=1-p$,
$\sigma_0\leftrightarrow\sigma_1$.
Indeed, let $Y=\sigma_0F(h)=\sigma_0f(\l)$ be the conductance
of the network $G$, shown in Figure~1.
Then the network $G'$,
obtained from $G$ by interchanging all the bond conductances
according to $\sigma_0\leftrightarrow\sigma_1$,
has a conductance $Y'=\sigma_1F(1/h)=\sigma_1f(1-\l)$.
For a large enough network, if $G$ is typical of the concentration $p$,
then $G'$ is typical of the concentration $1-p$.
The conductivity, the spectral function,
and the density of resonances therefore obey the following identities,
on any regular lattice
$$
\eqalignno{
\Sigma(p,h)&=h\Sigma(1-p,1/h),&(2.33\r{a})\cr
\l\Sigma(p,\l)&=(\l-1)\Sigma(1-p,1-\l),&(2.33\r{b})\cr
xH(p,x)&=(1-x)H(1-p,1-x),&(2.33\r{c})\cr
\rho(p,x)&=\rho(1-p,1-x).&(2.33\r{d})
}
$$

More interestingly, the square lattice is self-dual,
i.e., invariant under the geometric transformation called duality.
This concept has been introduced in physics by Kramers and Wannier~[10],
while its consequences on random resistor networks
have been explored in a systematic way by Straley~[11].
The dual $\w G$ of a planar network $G$ has bonds which cross those of $G$,
the conductances of any pair of crossing bonds being inverse to each other.
The duality property implies that the conductance of
the whole network $\w G$ reads $\w Y=1/Y=Z$.
With the same notations as above,
we have $\w Y=(1/\sigma_0)\w F(1/h)=(1/\sigma_0)\w f(\l)$,
so that $F(h)\w F(1/h)=f(\l)\w f(1-\l)=1$.
Therefore
$$
\Sigma(p,\l)\Sigma(p,1-\l)=\sigma_0^2,
\eqno(2.34)
$$
and
$$
\rho(p,x)=\rho(p,1-x).
\eqno(2.35)
$$
Because of its non-linearity, the duality identity~(2.34) for the conductivity
does not yield any identity involving the spectral function $H(p,x)$ only.

The above identities can be combined with the homogeneity properties~(2.33).
We thus obtain that the percolation threshold is $p_c=1/2$,
and that the conductivity and the spectral function right at this point read
$$
\Sigma(1/2,\l)=\sqrt{\sigma_0\sigma_1}=\sigma_0\sqrt\frac{\l-1}{\l},\qquad
H(1/2,x)=\frac{1}{\pi}\sqrt\frac{1-x}{x}.
\eqno(2.36)
$$
Furthermore, the spectral and total densities
of resonances obey the relations
$$
\rho(p,x)=\rho(p,1-x)=\rho(1-p,x)=\rho(1-p,1-x),\qquad\rho_R(p)=\rho_R(1-p).
\eqno(2.37)
$$

\medskip
\noindent {\bf 2.6 Critical behaviour and scaling}
\smallskip

As mentioned in the Introduction, the conductivity exhibits scaling behaviour
around the percolation threshold $p=p_c$.
The conductivity of the conductor-insulator mixture vanishes for $p\le p_c$,
while it reads
$$
\Sigma(\sigma_1=0)=\sigma_0A(p)\qquad(p>p_c).
\eqno(2.38)
$$
Similarly, the conductivity of the superconductor-conductor mixture
is infinite for $p\ge p_c$, while it reads
$$
\Sigma(\sigma_0=\infty)=\sigma_1B(p)\qquad(p<p_c).
\eqno(2.39)
$$
Both amplitudes $A(p)$ and $B(p)$ have a power-law behaviour
for $p$ near $p_c$:
$$
\matrix{
A(p)\approx a(p-p_c)^t\hfill& &(p\to p_c^+),\cr
B(p)\approx b(p_c-p)^{-s}\hfill& &(p\to p_c^-).
}
\eqno(2.40)
$$

The conductivity of the binary mixture obeys a scaling law
in the critical region defined by $\abs{h}\ll 1$ and $\abs{p-p_c}\ll 1$,
of the form~[1]
$$
\Sigma(p,h)\approx\sigma_0
\abs{p-p_c}^t\,\Phi_\pm\Bigl(h\abs{p-p_c}^{-s-t}\Bigr),
\eqno(2.41)
$$
where the $\Phi_\pm$ are scaling functions of one complex variable,
with $\pm$ referring to the sign of $p-p_c$.
The homogeneity relation~(2.33a) allows to describe the vicinity
of the other critical point, $\abs{h}\gg 1$ and $\abs{q-p_c}\ll 1$.

The scaling formula~(2.41) reproduces the power laws~(2.40)
for small values of the argument of the scaling functions,
as we have $a=\Phi_+(0)$, while $\Phi_-(x)\approx bx$ as $x\to 0$.
On the other hand, both scaling functions have the common
power-law behaviour $\Phi_\pm(x)\approx Kx^u$, with $u=t/(s+t)$,
as $\abs{x}\to\infty$ and $\abs{\arg x}<\pi$, hence
$$
\Sigma(p_c,h)\approx K\sigma_0\,h^u
\approx K\bigl(\sigma_0^s\sigma_1^t\bigr)^{1/(s+t)}
\eqno(2.42)
$$
for $\abs{\sigma_1}\ll\abs{\sigma_0}$, right at the percolation threshold.
More generally, the power-law behaviour~(2.42)
holds for $h_\star\ll\abs{h}\ll 1$,
where the crossover scale $h_\star$ reads
$$
h_\star\sim\abs{p-p_c}^{s+t},
\eqno(2.43)
$$
while the scaling laws~(2.38--40) hold in the opposite regime
$(\abs{h}\ll h_\star)$.

As a consequence of eq.~(2.28), the spectral function
also obeys a scaling law of the form~(2.41), namely
$$
H(p,x)\approx\abs{p-p_c}^t\,F_\pm\Bigl((1-x)\abs{p-p_c}^{-s-t}\Bigr),
\eqno(2.44)
$$
for $1-x\ll 1$ and $\abs{p-p_c}\ll 1$,
and a similar law around the other critical point,
i.e., for $x\ll 1$ and $\abs{q-p_c}\ll 1$.
We shall come back to the scaling law~(2.44) in section 2.9.

Consider now the binary model on a large but finite sample,
of size $M\times N$.
In the critical region, its mean conductance obeys the finite-size scaling law
$$
Y\approx\sigma_0N^{-t/\nu}
\,\Psi\Bigl((p-p_c)N^{1/\nu},(p-p_c)h^{-1/(s+t)},M/N\Bigr),
\eqno(2.45)
$$
where $\Psi$ is a three-variable scaling function.
As a consequence, right at the percolation threshold,
the critical region extends over a range
$$
\delta h\sim N^{-(s+t)/\nu}.
\eqno(2.46)
$$

On the square lattice, the duality symmetry implies
$p_c=1/2$, and $A(p)B(1-p)=1$, hence $s=t$ and $u=1/2$,
in agreement with eq.~(2.36).
The common numerical value of these exponents is
$s/\nu=t/\nu=0.9745\pm 0.0015$~[12],
with the exponent of the correlation length being exactly $\nu=4/3$,
hence $s=t=1.300$.
Hence, for $p=p_c=1/2$, the same critical singularity simultaneously affects
both endpoints of the spectrum, $\l=1$ and $\l=0$,
corresponding respectively to $h=0$ and $h=\infty$,
again in agreement with eq.~(2.36).

\medskip
\noindent {\bf 2.7 Sum rules}
\smallskip

The representation~(2.25) of the conductivity can be expanded
as the following series in inverse powers of $\l$
$$
\Sigma(p,\l)=\sigma_0\left(1-\sum_{k=0}^\infty\frac{H_k(p)}{\l^{k+1}}\right),
\eqno(2.47)
$$
where the coefficients
$$
H_k(p)=\int_0^1 x^kH(p,x)\d x
=\lim_{M,N\to\infty}\sum_{a=1}^{n_R}\alpha_a\l_a^k
\eqno(2.48)
$$
are the moments of the spectral function $H(p,x)$.

As we have already noticed in section 2.1, the expansion variable $1/\l$
is the dimensionless contrast between the conductances of both phases.
As a consequence, the expansion~(2.47) can be viewed as a special case of the
weak-disorder expansion of the conductivity of a random network
with an arbitrary narrow distribution of bond conductances~[13].
The general results to sixth order derived there can be transcribed
in the present case of binary disorder on the square lattice.
We thus obtain the following expressions for the first six moments,
$$
\eqalign{
&H_0(p)=q,\qquad H_1(p)=\frac{pq}{2},\qquad H_2(p)=\frac{pq}{4},\cr
&H_3(p)=\frac{pq}{8}(1+pq),\qquad
H_4(p)=\frac{pq}{16}\bigl(1+3pq+(J-1)pq(p-q)\bigr),\cr
&H_5(p)=\frac{pq}{32}\bigl(1+(J+5)pq+4(J-1)pq(p-q)-2(2J-3)p^2q^2\bigr),
}
\eqno(2.49)
$$
with $q=1-p$.
The number $J=J_1=J_2=1.092958179$, with the notations of ref.~[13],
is the only non-trivial quantity occurring in the sixth-order expansion.
We have thus derived explicit sum rules for the spectral function,
which agree with the expressions given in ref.~[7].
The number $J$, denoted there as $1-a_5^4$,
was estimated there from numerical data to be $J=0.9\pm0.5$.

\medskip
\noindent {\bf 2.8 Effective-medium approximation}
\smallskip

The effective-medium approximation (EMA), introduced by Bruggeman in 1935~[14],
is a self-consistent approximate scheme
to evaluate the conductivity of random impedance networks~[15, 16, 13],
which is still being very widely used~[17].

In the present case of the binary model on the square lattice,
the EMA prediction for the conductivity is given by
$$
p\,\frac{\Sigma^\EMA-\sigma_0}{\Sigma^\EMA+\sigma_0}
+(1-p)\frac{\Sigma^\EMA-\sigma_1}{\Sigma^\EMA+\sigma_1}=0,
\eqno(2.50)
$$
hence
$$
\eqalign{
\Sigma^\EMA&=\sigma_0\Bigl((p-1/2)(1-h)+\sqrt{(p-1/2)^2(1-h)^2+h}\Bigr)\cr
&=\frac{\sigma_0}{\l}\Bigl(p-1/2+\sqrt{(p-1/2)^2+\l(\l-1)}\Bigr).
}
\eqno(2.51)
$$
This EMA formula for the conductivity is analytic
in the $\l$-plane cut along the interval $\l_\min\le\l\le\l_\max$, with
$$
\l_\min=1/2-\sqrt{p(1-p)},\qquad
\l_\max=1-\l_\min=1/2+\sqrt{p(1-p)}.
\eqno(2.52)
$$
The prediction for the associated spectral function~[cf. eq.~(2.28)] reads
$$
H^\EMA(p,x)=\frac{\sqrt{x(1-x)-(p-1/2)^2}}{\pi x}
=\frac{\sqrt{(\l_\max-x)(x-\l_\min)}}{\pi x}.
\eqno(2.53)
$$

The EMA formula gives a very accurate approximation to the conductivity
of the binary model in generic circumstances.
For instance, the $1/\l$-expansion of the EMA formula~(2.51)
gives expressions $H^\EMA_k(p)$ for the moments which only differ from
the true results~(2.49), starting with $H_4(p)$,
by replacing $J$ by $J^\EMA=1$~[13].
The EMA scheme also respects the duality symmetry~(2.34).
The predictions~(2.51), (2.53) therefore agree with the exact results~(2.36)
for the conductivity and the spectral function at $p=p_c=1/2$.
The EMA also correctly predicts the equality $s=t$,
but not the common value of these exponents $(s=t=1$ instead of $1.300)$.

For a generic value of the concentration $(p\ne p_c)$,
the endpoints~(2.52) are such that $0<\l_\min<\l_\max<1$.
The support of the EMA prediction~(2.53) for the spectral function
thus does not extend over the whole interval $0\le\l\le1$.
In the vicinity of the percolation threshold $(p\to p_c=1/2)$,
we have $\l_\min=1-\l_\max\approx(p-p_c)^2$,
in agreement with the estimate~(2.43) of the crossover scale $h_\star$,
with $s+t=2$.

\medskip
\noindent {\bf 2.9 Lifshitz tails}
\smallskip

It has been argued~[6] that the spectral function $H(p,x)$
of the true conductivity of the binary model
extends over the whole allowed spectrum $0\le\l\le1$,
for any value of the concentration $p$, at variance with the EMA
formula~(2.53).

This argument has then been put in perspective~[1]
with Lifshitz singularities~[18--20].
These singularities are caused by the presence of
very large, and thus very improbable,
ordered regions in a randomly disordered system.
The original example considered by Lifshitz~[18] is that of the phonon spectrum
of a binary harmonic alloy, consisting of light atoms,
with mass $m$ and concentration $p$,
and heavy atoms, with mass $M>m$ and concentration $q=1-p$.
Lifshitz has argued that the vicinity of the upper edge
$\omega_\max$ of the phonon spectrum of the alloy is dominated
by large ordered regions, almost spherical in shape,
consisting only of light atoms.
He thus showed that $\omega_\max$ coincides with
the upper edge of the pure lattice consisting only of light atoms,
and that the density of states of the alloy vanishes
exponentially fast near $\omega_\max$, as
$$
\rho(\omega)\sim\exp\left(-c\abs{\ln p}(\omega_\max-\omega)^{-d/2}\right),
\eqno(2.54)
$$
where $c$ is a lattice-dependent constant, which can be evaluated exactly.
Along this line of thought, it has been argued in ref.~[1]
that the spectral function of the binary model
has an exponentially small Lifshitz tail,
extending all the way to the endpoints $\l=0$ and $\l=1$
of the spectrum of resonances, of the form
$$
H(p,x)\sim\exp\left(-C(p)x^{-d/2}\right)\qquad(x\to0),
\eqno(2.55)
$$
and similarly for $x\to1$.
This expression was rather conjectural,
as the determination of the relevant ordered regions was left
as an open question, so that the prefactor $C(p)$ was not predicted.

Hesselbo~[21] then argued
that the relevant ordered regions are hairpin configurations,
as shown in Figure~2a.
Let $Y_n$ be the transversal conductance of the hairpin
consisting of $n$ cells,
measured between the point electrodes shown in Figure~2a,
considered as an isolated network (not embedded in the square lattice).
This quantity can be evaluated from the recursion relation
$$
Y_n=\sigma_1+\left(\frac{2}{\sigma_0}+\frac{1}{Y_{n-1}}\right)^{-1},
\eqno(2.56)
$$
with $Y_0=\sigma_0$.
Consider first a value of $h$ not on the negative real axis, and set
$$
h=\frac{\sigma_1}{\sigma_0}=2\sinh^2\mu,
\eqno(2.57)
$$
with $\re\mu>0$ and $\abs{\im\mu}<\pi$.
The M\"obius map involved in the recursion~(2.56) has two fixed points,
$$
Y_\pm=\frac{\sigma_0}{2}\bigl(e^{\pm2\mu}-1\bigr),
\eqno(2.58)
$$
with the stable fixed point $Y_+$ being the transversal
conductance of the infinitely long hairpin (ladder).
The recursion~(2.56) can be solved explicitly, along the lines of ref.~[1],
by means of the variable $t_n=(Y_n-Y_-)/(Y_n-Y_+)$.
We thus obtain
$$
Y_n=\sigma_0\sinh\mu
\,\frac{3\cosh((2n+1)\mu)-\cosh((2n-1)\mu)}{3\sinh(2n\mu)-\sinh(2(n-1)\mu)}.
\eqno(2.59)
$$

For a strong dielectric contrast ($h\to 0$),
the correlation length of currents along the hairpin
diverges according to $\xi=1/(2\mu)\approx(2h)^{-1/2}$.
In this regime, the conductance of long hairpins scales as
$$
Y_n\approx\frac{\sigma_0}{2n}\,z\,\r{cotanh}\,z,\qquad\hbox{with}\quad
z=\frac{n}{\xi}=2n\mu.
\eqno(2.60)
$$
This scaling form of the conductance exhibits an infinite array of
alternating zeros, lying at
$$
\w z_a=\frac{(2a+1)i\pi}{2},\qquad\hbox{i.e.,}\quad
1-\w\l_a\approx\frac{(2a+1)^2\pi^2}{8n^2},
\eqno(2.61)
$$
and poles, lying at
$$
z_a=ai\pi,\qquad\hbox{i.e.,}\quad 1-\l_a\approx\frac{a^2\pi^2}{2n^2},
\eqno(2.62)
$$
with $a\ge1$.
The first zero $\w\l_1$, corresponding to a pole in the dual configuration,
yields, according to Hesselbo, the Lifshitz behaviour of the spectral function,
at least for a small enough concentration $p$.

It turns out that the formula~(2.59) also gives the conductance
of worm-like networks, such as the configuration shown in Figure~2b,
where the square cells are put together in any random fashion,
respecting the linear structure and the constraint of self-avoidance.
The number of such worm-like configurations
with $n$ cells is of order $\exp(nS)$,
with $S$ being the associated configurational entropy.
On the other hand, a hairpin with $n$ cells
occurs with a probability of order $p^{2n}$, at least for $p$ small enough.
Altogether, the first zero of eq.~(2.61) is expected to show up
with a probability weight of order $(p^2e^S)^n$.
By eliminating the number $n$ between the above estimates,
we obtain the following analytical form for the Lifshitz tail
of the density of resonances and of the spectral function
$$
\rho(p,x)\sim H(p,x)\sim\exp\left(-\frac{C(p)}{\sqrt{x}}\right)\qquad(x\to0),
\eqno(2.63)
$$
and a similar formula for $x\to1$.
This result, with an inverse-square-root behaviour
in the spectral variable $x$,
is characteristic of Lifshitz tails in one-dimensional systems~[19].
This is due to the fact that the relevant structures are linear objects.
Furthermore, the identities~(2.33), (2.37) imply $C(p)=C(1-p)$.
The above argument also leads to a prediction for the
small-$p$ behaviour of the amplitude $C(p)$, namely
$$
C(p)\approx\frac{\pi}{\sqrt2}\left(\abs{\ln p}-\frac{S}{2}\right)
\qquad(p\ll 1).
\eqno(2.64)
$$
These predictions will be compared to numerical data in section 4.

The Lifshitz tail of the spectral function manifests itself
in the long-time tail of the transient intensity response of the $R-C$ model.
Indeed, in the regime of long times $(t\gg\tau)$,
eq.~(2.30) is dominated by the vicinity of $x=0$, hence
$$
I(t)
\sim\int_0^\infty\d x\exp\left(-x\frac{t}{\tau}-\frac{C(p)}{\sqrt{x}}\right)
\sim\exp\left\{-3\left(\frac{C(p)^2}{4}\frac{t}{\tau}\right)^{1/3}\right\}.
\eqno(2.65)
$$

In the critical region, the prediction~(2.63) for the Lifshitz tail
is compatible with the scaling law~(2.44) for the spectral function,
provided the amplitude $C(p)$ vanishes near the percolation threshold,
according to
$$
C(p)\sim\abs{p-p_c}^{(s+t)/2}.
\eqno(2.66)
$$
The Lifshitz behaviour~(2.63) is expected to hold only deep in the tails,
for $x$ (or $1-x$) much smaller than the crossover scale $h_\star$,
defined in eq.~(2.43).

\medskip
\noindent {\bf 3 ALGORITHM}
\smallskip

We now turn to the presentation of our algorithm
for evaluating the rational $h$-dependence of the conductance
of a finite binary network, such as that shown in Figure~1.
This approach is an extension of the method of ref.~[2].
It turns out that a very similar formalism
was proposed by Straley~[6] some twenty years ago,
but this work has apparently not been noticed since then.
The conductance will be determined in the second form of eq.~(2.1a),
namely we shall calculate first the poles $\l_a$ of the conductance,
giving the positions of the resonances,
and then the associated residues $\alpha_a$,
giving the strengths of the resonances.

\medskip
\noindent {\bf 3.1 Generalities}
\smallskip

Along the lines of ref.~[2],
our starting point is the Kirchhoff equations
for $V_\x$, the electric potential at site $\x$:
$$
\sum_{\y(\x)}\sigma_{\x,\y}(V_\x-V_\y)=0.
\eqno(3.1)
$$
There is one such equation per site $\x$ inside the network.
The notation $\y(\x)$ means that $\y$ is a neighbour of $\x$,
i.e., there exists a bond $(\x,\y)$,
and the sum possibly includes sites $\y$ belonging to either electrode.
Eqs.~(3.1) have to be complemented by the boundary conditions
$V_\y=0$ for the sites $\y$ on the left electrode,
and $V_\y=V_0$ for the sites $\y$ on the right electrode.

We define the topological Laplace operator $\Delta$ on the network as
$$
(\Delta V)_\x=\sum_{\y(\x)}(V_\y-V_\x),
\eqno(3.2)
$$
again with the convention that the sum possibly includes
sites $\y$ belonging to either electrode,
in which case the Dirichlet boundary condition $V_\y=0$ is assumed.
The operator $\Delta$ can be written as the sum $\Delta=\Delta_\P+\Delta_\Q$
of its components $\Delta_\P$ on the $\P$-set and $\Delta_\Q$ on the $\Q$-set,
respectively defined as
$$
(\Delta_\P V)_\x=\sum_{\y\in\P(\x)}(V_\y-V_\x),\qquad
(\Delta_\Q V)_\x=\sum_{\y\in\Q(\x)}(V_\y-V_\x),
\eqno(3.3)
$$
where $\y\in\P(\x)$ (respectively, $\y\in\Q(\x)$)
means that $(\x,\y)$ is a $\P$-bond (respectively, a $\Q$-bond).
We also introduce the quantities
$$
\left\{\matrix{
\hbox{$A\L_\x=1$ iff $\x$ is a neighbour of the left electrode,}\hfill\cr
\hbox{$A\R_\x=1$ iff $\x$ is a neighbour of the right electrode,}\hfill\cr
\hbox{$B\L_\x=1$ iff $\x$ is connected to the left electrode
by a $\Q$-bond,}\hfill\cr
\hbox{$B\R_\x=1$ iff $\x$ is connected to the right electrode
by a $\Q$-bond.}\hfill\cr
}\right.
\eqno(3.4)
$$

With these notations, the Kirchhoff equations~(3.1) read
$$
\sigma_0(\Delta_\P V)_\x+\sigma_1(\Delta_\Q V)_\x
+V_0\Bigl(\sigma_0\bigl(A\R_\x-B\R_\x\bigr)+\sigma_1B\R_\x\Bigr)=0,
\eqno(3.5)
$$
or equivalently, in vector and matrix notation,
$$
(\Delta_\Q-\l\Delta)\V=V_0\bigl(\l\A\R-\B\R\bigr).
\eqno(3.6)
$$
This reduced form only involves the complex variable $\l$ defined in eq.~(1.3).
The conductance of the network is given by
$$
Y=\frac{I}{V_0},
\eqno(3.7)
$$
where the total current $I$
flowing into the network from the left electrode reads
$$
I=\sum_\x\Bigl(\sigma_0\bigl(A\L_\x-B\L_\x\bigr)+\sigma_1B\L_\x\Bigr)V_\x,
\eqno(3.8)
$$
or equivalently, in vector notation,
$$
I=\frac{\sigma_0}{\l}\bigl(\l\A\L-\B\L\bigr)\cdot\V.
\eqno(3.9)
$$

\medskip
\noindent {\bf 3.2 Poles of the conductance}
\smallskip

In analogy with ref.~[2],
the poles of the conductance are the non-trivial values $\l_a$ of $\l$
for which the homogeneous Kirchhoff equations~(3.1) with $V_0=0$
have a non-zero solution, namely
$$
(\Delta_\Q-\l_a\Delta)\V=\0.
\eqno(3.10)
$$
This is a well-posed generalised eigenvalue problem,
since $\Delta$ and $\Delta_\Q$ are two real symmetric matrices,
of size $n_S\times n_S$, and $(-\Delta)$ is a positive definite matrix.

It turns out that the endpoints $\l=0$ or $\l=1$ are in general
extensively degenerate eigenvalues of eq.~(3.10).
These eigenvalues do not correspond to resonances.
Indeed, we know from section 2.1 that the conductance has no pole at $\l=1$,
while it has a simple pole at $\l=0$ if, and only if,
the $\Q$-phase is conducting.
Let us set $\eps_\Q=1$ in this situation, $\eps_\Q=0$ else.
The number of resonances then reads
$$
n_R=n_S-n_0-n_1+\eps_\Q,
\eqno(3.11)
$$
where $n_0$ and $n_1$ denote the respective multiplicities
of the endpoint eigenvalues $\l=0$ and $\l=1$.
More precisely, $n_0$ is the number
of zero-modes of the operator $\Delta_\Q$, i.e., the dimension of its kernel.
Since $\Delta_\Q$ has one zero-mode per cluster $C$
of the $\Q$-phase which is disconnected from the electrodes, we have
$$
n_0=n_S-\sum_{C\subset\Q}\bigl(s(C)-1+\chi(C)\bigr).
\eqno(3.12)
$$
In this formula, the sum runs over the clusters $C$ of the $\Q$-phase,
$s(C)$ denotes the number of sites of the cluster $C$,
and the characteristic function $\chi(C)$ is unity if the cluster $C$
overlaps with either of the electrodes, and zero else.
The multiplicity $n_1$ of the eigenvalue $\l=1$
can be expressed similarly,
in terms of the clusters of the $\P$-phase.

In the thermodynamic limit, one can derive from eq.~(3.12)
expressions for the fractions of eigenvalues which are condensed
at $\l=0$ and $\l=1$.
Indeed, the terms $\chi(C)$ are negligible,
while the other terms can be expressed as functions
of geometrical characteristics of the bond percolation problem~[22], namely
$$
\eqalign{
\rho_0(p)&=\lim_{M,N\to\infty}\frac{n_0}{MN}=1-2q+P(q)+n_c(q),\cr
\rho_1(p)&=\lim_{M,N\to\infty}\frac{n_1}{MN}=1-2p+P(p)+n_c(p),
}
\eqno(3.13)
$$
where $n_c(p)$ is the mean number of finite clusters per site,
while $P(p)$, the percolation probability,
is the probability for any given bond to belong to the infinite cluster.
The latter quantity is non-vanishing only for $p>p_c$.

The density of resonances $\rho_R(p)$, defined in eq.~(2.32), then reads
$$
\rho_R(p)=1-\rho_0(p)-\rho_1(p)=1-P(p)-P(q)-n_c(p)-n_c(q).
\eqno(3.14)
$$
This quantity is symmetric in the exchange $p\leftrightarrow q$,
in agreement with eq.~(2.37).
For a small concentration $p$,
the contributions to eq.~(3.12) of all the $\P$-clusters and $\Q$-clusters
consisting of up to four bonds can be enumerated by hand.
We thus obtain
$$
\rho_0(p)=p^4+\cdots,\qquad
\rho_1(p)=1-2p+p^4+\cdots,\qquad
\rho_R(p)=2p-2p^4+\cdots
\eqno(3.15)
$$
At the percolation threshold, $\rho_R(p)$ takes its maximal value,
which can be determined as follows.
The percolation probability $P(p_c)$ vanishes,
while $n_c(p_c)$ is known exactly~[23], hence
$$
\rho_0(p_c)=\rho_1(p_c)=n_c(p_c)=\frac{3\sqrt{3}-5}{2}=0.098076,\qquad
\rho_R(p_c)=3(2-\sqrt{3})=0.803848.
\eqno(3.16)
$$

\medskip
\noindent {\bf 3.3 Residues of the conductance}
\smallskip

At any resonance corresponding to a non-trivial eigenvalue
$(\l_a\ne0$ and 1) of eq.~(3.10),
the map of the electric potentials on the network
is given by the components $(X_a)_\x$ of the associated
right eigenvector $\X_a$.
Since eq.~(3.10) is symmetric, the $\X_a$ are simultaneously its left
and its right eigenvectors:
$$
\X_a^t(\Delta_\Q-\l_a\Delta)=\0,\qquad(\Delta_\Q-\l_a\Delta)\X_a=\0,
\eqno(3.17)
$$
with the row vector $\X^t$ being the transposed of the column vector $\X$.
The $n_S$ eigenvectors $\{\X_a\}$ form a basis.
They are orthogonal to each other with respect to the metric $(-\Delta)$,
namely $\X_a^t(-\Delta)\X_b=0$ for $a\ne b$.
We normalise them as
$$
\X_a^t(-\Delta)\X_b=\delta_{a,b}.
\eqno(3.18)
$$
The squared norm of the eigenvector $\X_a$ reads
$$
\X_a^t(-\Delta)\X_a=\sum_{(\x,\y)}E_{\x,\y}^2=1,
\eqno(3.19)
$$
where
$$
E_{\x,\y}=(X_a)_\x-(X_a)_\y
\eqno(3.20)
$$
is the local electric field on the bond $(\x,\y)$.

The solution $\V$ to the inhomogeneous Kirchhoff equations~(3.1), (3.6)
can be obtained in terms of the eigenvectors $\X_a$.
Indeed, let us expand $\V$ on the basis of the $\X_a$:
$$
\V=\sum_{a=1}^{n_S}c_a(\l)\X_a.
\eqno(3.21)
$$
By inserting this expansion into eq.~(3.6),
and multiplying to the left by $\X_b^t$,
we readily obtain the amplitude $c_b(\l)$ in the form
$$
c_b(\l)=V_0\frac{\bigl(\l\A\R-\B\R\bigr)\cdot\X_b}{\l-\l_b}.
\eqno(3.22)
$$
Finally, by inserting this result into eqs.~(3.9), (3.7),
we obtain the following formula for the residues $\alpha_a\ge0$
of the conductance at its non-trivial poles $(\l_a\ne0$ and 1)
$$
\alpha_a=-\frac{M}{N\l_a}
\Bigl(\bigl(\l_a\A\L-\B\L\bigr)\cdot\X_a\Bigr)
\Bigl(\bigl(\l_a\A\R-\B\R\bigr)\cdot\X_a\Bigr).
\eqno(3.23)
$$

As recalled in the beginning of section~3.1,
the conductance also exhibits a simple pole at $\l=0$,
if the $\Q$-phase is conducting.
The corresponding residue $\alpha_1$,
which yields the conductance of the $\Q$-phase by means of eq.~(2.3),
is not directly given by eq.~(3.23).
It can, however, be determined from the other ones
$(\alpha_2,\cdots,\alpha_{n_R})$, by using the sum rule~(2.5).

\medskip
\noindent {\bf 4 NUMERICAL RESULTS}
\smallskip

We have shown in section 3 that the full rational $\l$-dependence
of the conductance of a finite binary network
can be expressed in terms of the generalised eigenvalues
and eigenvectors of eq.~(3.10).
We have implemented this algorithm numerically, using the IMSL routine EIGZS,
in order to obtain numerical data concerning several quantities of interest,
which will be discuss successively throughout this section.
The CPU time for solving the spectral problem for each sample
grows rapidly with the system size, proportionally to $n_S^3$,
i.e., to $N^6$ for a square sample of size $N\times N$.

The optimal use of our algorithm therefore consists in obtaining good
statistics on samples of moderate sizes.
We have commonly used sample sizes such as $N=16$ or $N=20$,
and a statistical ensemble of several times $10^4$ samples,
having some $10^7$ random bond conductances in total,
a good enough statistics to obtain very accurate data.
No observable systematic finite-size effects have been found,
even in the critical regions $(p\to p_c$ and $\l\to 0$ or $\l\to 1)$.
This observation is in agreement with the argument,
given at the end of section 4.1,
showing that the critical regions are indeed very small in the variable $\l$.

\vfill\eject
\noindent {\bf 4.1 Density of resonances and spectral function}
\smallskip

We have evaluated the spectral density of resonances $\rho(p,x)$
and the spectral function $H(p,x)$,
by means of their respective definitions~(2.31) and (2.26).
The spectral function,
which plays a central role in the Bergman-Milton formalism~[4, 5],
had only been the subject of a limited amount of work.
In ref.~[7] it has been extracted from the imaginary part
of the conductance, measured at a small but finite distance
$\epsilon$ from the negative real $h$-axis.
The present method yields a direct measurement of the spectral function,
avoiding especially any contamination from the delta-function
at $x=0$ [see eq.~(2.27)], which can be discarded in an exact way.

Figures~3 and 4 respectively show histogram plots of $\rho(p,x)$ and $xH(p,x)$,
for values of the concentration $p$ ranging from 0.05 to 0.5.
Indeed, the symmetry relations~(2.33), (2.37) allow to restrict our attention
to $p\le p_c=1/2$.
Each plot contain the accumulated data of an ensemble of
over 20,000 configurations of a network of size $N=16\times 16$,
corresponding to $10^7$ random bonds in total.
The density of resonances $\rho(p,x)$ exhibits
the expected symmetry~(2.37) under the transformation $x\leftrightarrow1-x$
within a good accuracy, for all values of $p$.
This demonstrates that we have used a large enough statistical ensemble
of random networks.
The product $xH(p,x)$ does not possess such a symmetry
(except for $p=p_c=1/2$),
while the EMA prediction~(2.53), shown as a semi-circle in Figure~4,
is symmetric under $x\leftrightarrow1-x$.
Both $\rho(p,x)$ and $xH(p,x)$ exhibit a rich structure,
down to the scale of the resolution (each plot contains 100 bins).
It will be demonstrated more clearly in section 4.2
that they are non-vanishing over the whole allowed spectrum $(0\le x\le 1)$.

For a small enough concentration,
the most salient structures in $\rho(p,x)$ and $H(p,x)$
can be predicted from the analysis of the resonances
of isolated finite clusters.
For consistency with ref.~[2], we shall consider the regime $q=1-p\to 0$.
To leading order in $q$,
the relevant configuration consists of one isolated $\Q$-bond
embedded in a host lattice consisting of $\P$-bonds.
This one-bond cluster, shown as configuration A in Figure~5,
has one single resonance, at $\l=\l_{\r{A}1}=1/2$,
yielding the observed leading peak in $\rho(p,x)$ and $xH(p,x)$.
This one-bond cluster has two possible orientations,
but only the horizontal case yields a non-vanishing residue~[2],
to leading order as $q\to 0$.
The spectral function, the density of resonances,
and the mean number of resonances therefore behave as
$$
H(p,x)\approx p\,\delta(x-1/2),\qquad
\rho(p,x)\approx 2p\,\delta(x-1/2)\qquad(p\to 0),
\eqno(4.1)
$$
in agreement with the sum rules~(2.49) and with eq.~(3.15),
to leading order in $p$.
To second order in $p$, $\rho(p,x)$ and $H(p,x)$
consist of a countable infinity of discrete components (delta-functions),
corresponding to the resonances of configurations consisting
of two bonds, in arbitrary relative position and orientation~[2],
and so on.
The most salient subleading peaks
have been marked in Figures~3 and 4 (dashed verticals)
by some of the resonances of the two-bond
and three-bond configurations shown in Figure~5.
The configurations A, B, and E are self-dual,
while (C,D) and (F,G) form dual pairs.
The resonances of these clusters,
determined exactly along the lines of ref.~[2], are given in Table 1.

Right at the percolation threshold $(p=p_c=1/2)$,
the data for the spectral function (Figure~4e)
agree with the exact analytical result~(2.36),
which coincides with the EMA prediction.
The accuracy of this agreement provides another check of the quality
of the numerical simulations.
The density of resonances remains a non-trivial function
$\rho(p_c,x)$ at the percolation threshold (Figure~3e).
Integrating the data of this Figure
leads to the estimate $\rho_R(p_c)\approx 0.80$,
again in good agreement with the exact result~(3.16).
It is worth noticing that the data shown in Figures~3e
and 4e are practically not affected by the critical singularities
at the endpoints of the spectrum.
Indeed, the size of the critical region, given by eq.~(2.46),
can be estimated to be a few times $10^{-3}$,
i.e., smaller than the width of the first or last bin.

As the concentration varies from $p=0$ to $p=p_c=1/2$,
the profiles of the density of resonances
and of the spectral function deform in a progressive way.
They get smoother and smoother, with their maxima moving in a continuous way.
The dashed verticals are shown as guides for the eye on all plots
of Figures~3 and 4, although they only label the most salient structures
for a small enough concentration.
The spectral function also gets progressively in better and better agreement
with the EMA prediction.

\medskip
\noindent {\bf 4.2 Lifshitz tails}
\smallskip

We have also investigated the behaviour of the density of resonances
near the endpoints $x=0$ and $x=1$,
in order to check the prediction~(2.63) of the Lifshitz tail.
We have chosen to investigate the density of resonances
$\rho(p,x)$, rather than the spectral function,
because $\rho(p,x)$ can be expected on general grounds to exhibit
a clearer signal, in analogy with the one-dimensional situation~[19].
Furthermore, the statistics can be doubled
by using the symmetry relations~(2.37), and each sample requires less CPU time,
since the calculation of the eigenvectors $\X_a$ is not required for
$\rho(p,x)$.

Figure~6 shows a logarithmic plot of the integrated spectral
density of resonances
$$
\rho_\r{int}(p,x)=\int_0^x\rho(p,y)\d y
\eqno(4.2)
$$
against $x^{-1/2}$, for a concentration $p=0.1$.
The data correspond to over 103,000 samples of size $N=16\times 16$,
i.e., to $5\times 10^7$ random bonds.
The range of the plotted data corresponds to $x\le\l_\min$
and $\l_\max\le x\le1$,
with $\l_\min$ and $\l_\max$ being the endpoints~(2.52) of the EMA
prediction for the spectral function.
Indeed, the Lifshitz behaviour is expected to manifest itself mostly
out of the ``bulk'' of the spectrum, the latter being conveniently
defined as the support of the EMA formula~[20].
A linear behaviour is clearly observed,
confirming the analytical form~(2.63) of the Lifshitz tail.
A further qualitative confirmation of Hesselbo's argument
on the Lifshitz behaviour is as follows.
The data of Figure~6 exhibit oscillations around to the fitted straight line,
and the top of each of the most prominent of these oscillations corresponds,
with a good accuracy,
to the lowest resonance of the hairpin structures shown in Figure~2a,
embedded in the square lattice, with $n=1$ to 4 cells.
The case $n=1$ corresponds to configuration G of Figure~5,
with its lowest resonance $\l_{\r{G}1}$.

From a quantitative viewpoint,
the slope of the fitted straight line in Figure~6
yields $C(p=0.1)\approx 3.19$.
The amplitude $C(p)$ has been similarly measured for $p=0.05$, 0.15, and 0.2.
The results are plotted in Figure~7 against $\abs{\ln p}$.
The data are nicely fit to the straight line
$C(p)\approx 1.98\abs{\ln p}-1.39$,
to be compared with the analytical prediction~(2.64).
The slope 1.98 is some 10\% smaller
than the analytical value $\pi/\sqrt{2}=2.2214$.
The intercept yields the estimate $S\approx 1.40$
for the configurational entropy per cell,
a significantly larger value than the entropy of self-avoiding walks
on the square lattice, $S_{\rm SAW}=\ln\mu_{\rm SAW}=0.970$~[24].
These observations suggest that other types
of linear extended structures, besides the worm-like ones identified
in the framework of Hesselbo's argument,
may contribute to the Lifshitz behaviour of the conductivity.

\medskip
\noindent {\bf 4.3 Distribution of spacings between resonances}
\smallskip

The distribution of spacings between successive energy levels
has been extensively investigated in a variety of quantum systems,
ranging from nuclei to billiards~[25].
Generic spectra belong to three universality classes
of level spacing distributions, according to their symmetry properties,
in correspondence with the classical ensembles
of real symmetric, Hermitian, and symplectic random matrices~[26],
respectively called GOE, GUE, and GSE.

We have investigated the distribution of spacings between successive resonances
in the range
$$
p=p_c=1/2,\qquad 1/4\le x\le 3/4,
\eqno(4.3)
$$
where the spectral density of resonances is very flat,
i.e., very close to being a constant, $\overline{\rho}=0.658$ (see Figure~3e).
The range~(4.3) can be considered as fully generic,
although the concentration assumes its critical value $p_c$,
since the critical singularities only influence very small
regions around the endpoints $x=0$ or $x=1$, as explained in section~4.1.

For a finite network of size $N\times N$,
the mean spacing between two successive resonances
is approximately $\l_{a+1}-\l_a\approx 1/(\overline{\rho}N^2)$.
We thus define the reduced spacings
$$
s_a=\overline{\rho}N^2(\l_{a+1}-\l_a).
\eqno(4.4)
$$
Figure~8 shows a histogram plot
of the distribution $P(s)$ of the spacings $s_a$,
obtained from an ensemble of networks of size $20\times 20$
having $10^7$ random bonds in the range~(4.3) in total.
This distribution obeys by construction
$\int_0^\infty P(s)\d s=\int_0^\infty sP(s)\d s=1$.

Since the generalised eigenvalue problem~(3.10)
involves two real symmetric matrices, $\Delta$ and $\Delta_\Q$,
the natural universality class to which the data for $P(s)$ should be compared
is the Gaussian orthogonal ensemble (GOE).
The law $P_\r{GOE}(s)$~[26] is shown as a full line in Figure~8.
The data share characteristic qualitative features
of the GOE spacing distribution,
including a linear repulsion at short spacings,
i.e., $P(s)\sim s$ for $s\ll1$, and a fast fall-off at large spacings.
There is, however, a small but significant quantitative
difference between the data in the range~(4.3) and the GOE prediction.

\medskip
\noindent {\bf 4.4 Number of visible resonances}
\smallskip

We now turn to the number of visible resonances of the $RL-C$ model
on a finite network.
A resonance is said to be visible if it corresponds to a true maximum
in the frequency dependence of the real part
of the admittance, as given by eq.~(2.20a).

For a finite network of size $N\times N$,
the typical spacing between resonances scales as $1/N^2$,
as already mentioned in section 4.3.
Since two resonances of comparable strengths are resolved,
i.e., separately visible,
if their spacing is larger than the width $1/Q$ of each of them,
we are led to propose the following finite-size scaling law
for the fraction of visible resonances
$$
\frac{(n_R)_\r{visib}}{n_R}\approx F\left(\frac{Q}{N^2}\right).
\eqno(4.5)
$$
The scaling function $F(X)$ is expected to grow monotonically
from $F(0)=0$, since a vanishing fraction of the resonances is visible
if $Q$ does not scale as $N^2$, to $F(\infty)=1$,
since all the resonances of a finite sample are eventually visible
for a large enough quality factor.

Figure~9 shows a plot of the fraction of visible resonances
over the whole spectrum $(0\le\l\le1)$, for $p=p_c=1/2$.
The observed collapse of the data for the sizes $12\times 12$ and $20\times 20$
clearly demonstrates the validity of the scaling law~(4.5).
The full line shows a numerical fit of both series of data to the
common analytical form $1/\bigl(1-F(X)\bigr)^2=1+XP_2(X)$,
with $P_2(X)$ being a quadratic polynomial.
The quality of the fitted curve, meant as a guide to the eye,
suggests a linear behaviour of the scaling function at small $X$,
as well as a power-law convergence of the form
$$
1-F(X)\sim X^{-3/2}\qquad(X\gg 1).
\eqno(4.6)
$$

The fraction of visible resonances depends a priori
on how uniformly the resonances are distributed,
and on the dispersion in the corresponding residues $\alpha_a$.
Both features can depend quantitatively on the concentration $p$.
The scaling function $F(X)$ is therefore expected not to be universal,
but rather to weakly depend on the concentration $p$
and on the range of values of $\l$ under consideration.
Its main qualitative features, such as the power law~(4.6),
are however expected to be universal.
The same remarks apply to the distribution of spacings
between resonances, investigated in section 4.3.

\medskip
\noindent {\bf 4.5 Distribution of local electric fields}
\smallskip

The algorithm presented in section 3 also gives access to the spatial structure
of the resonances.
Indeed, the eigenvector $\X_a$
directly provides a map of the electric potentials
at the resonance corresponding to the eigenvalue $\l_a$.
For each resonance, we define the local electric field on the bond $(\x,\y)$ as
$$
E_{\x,\y}=V_\x-V_\y=c\bigl((\X_a)_\x-(\X_a)_\y\bigr).
\eqno(4.7)
$$
The electric fields are defined up to an overall multiplicative constant $c$.
We choose this constant to be $c=\sqrt{n_B}$, so that the normalisation~(3.19)
of the eigenvectors implies
$$
\sum_{(\x,\y)}E_{\x,\y}^2=n_B.
\eqno(4.8)
$$

One could think of many ways of analysing the spatial structure
of the electric field at resonance.
We have chosen to investigate the distribution of the local field
living on any given bond.
More precisely, we have evaluated the successive moments of this distribution
on square samples of size $N\times N$, namely
$$
S_k(N)=\left\langle\abs{E}^k\right\rangle
=\left\langle\frac{1}{n_B}\sum_{(\x,\y)}\abs{E_{\x,\y}}^k\right\rangle.
\eqno(4.9)
$$
The normalisation~(4.8) ensures the identities $S_0(N)=S_2(N)=1$.

Figure~10 shows a log-log plot of the first non-trivial moments,
of index $k=1$, 3, 4, 5, and 6,
against the linear size $N$ of the sample, from $N=6$ to $N=24$.
Data are obtained in the range~(4.3),
with around $10^7$ random bonds for each sample size.
Power laws of the form
$$
S_k(N)\sim N^{x_k}
\eqno(4.10)
$$
are clearly apparent.
This scaling behaviour, with a non-trivial dependence of the exponent $x_k$
on the index $k$, is a signature of multifractality~[27, 28].
Along the lines of the multifractal formalism,
we introduce the generalised (R\'enyi) dimensions $d_k$, such that
$$
x_k=\frac{(k-2)(2-d_k)}{2}.
\eqno(4.11)
$$
The $d_k$ are expected to decrease from $d_0=2$, the dimension of the network,
to $d_\infty=0$.

In physical terms, multifractality implies that
the patterns of resonant electric fields exhibit strong local fluctuations,
rather similar to those observed in wavefunctions of the Anderson model,
in the marginal two-dimensional situation~[29].
In particular the resonances are neither localised nor extended,
in the conventional sense of these expressions.
Indeed, extended patterns of electric fields
would correspond to $d_k=2$ for all $k\ge0$,
while localised ones would have $d_k=0$ for $k\ge1$.
We recall that a similar phenomenon of multifractality
has been reported for the DC problem of the conductor-insulator mixture,
right at the percolation threshold~[30, 31].

From a quantitative viewpoint, a more refined fit of the data to the
power laws~(4.10), including a relative correction in $1/N$,
yields more accurate estimates for the exponents $x_k$
and the associated dimensions $d_k$, listed in Table 2.
Figure~11 shows a plot of the $d_k$ against the index $k$.
An approximate linear decay of the form
$$
d_k\approx 2-\beta k,
\eqno(4.12)
$$
with $\beta\approx 0.194$,
is observed over a fairly broad range $(0\le k\le 4)$.
A similar linear behaviour has been predicted analytically
for the two-dimensional Anderson model in the weak-disorder regime~[29].
The linear law~(4.12) corresponds to a log-normal (LN) distribution
of the local fields.
Indeed, let us set
$$
\ell=\ln\abs{E}.
\eqno(4.13)
$$
The scaling law~(4.10) then reads
$\langle\exp(k\ell)\rangle\sim\exp(Bk(k-2)/2)$, with $B=\beta\ln N$.
Neglecting the $k$-de\-pen\-den\-ce of the prefactor,
this last expression corresponds to a log-normal distribution for $\abs{E}$,
i.e., to a Gaussian law for the logarithmic variable $\ell$, namely
$$
\Pi\LN(\ell)=(2\pi B)^{-1/2}\exp\left(-\frac{(\ell+B)^2}{2B}\right).
\eqno(4.14)
$$

The actual probability density $\Pi(\ell)$ is shown in Figure~12,
for networks of size $16\times 16$ in the range~(4.3).
This very asymmetric distribution looks quite different from a log-normal law.
In particular, it falls off as $\Pi(\ell)\sim\exp(-\ell)$ as $\ell\to-\infty$,
yielding a finite value at $\abs{E}=0$
of the probability density $P(\abs{E})=e^\ell\Pi(\ell)$.
Most of the dependence of the distribution on the sample size $N$
takes place to the right of the plot, for large values of $\ell$,
close to the upper bound $\ell_\max=(\ln n_B)/2$,
where a very fast decay is observed.

\medskip
\noindent {\bf 5 DISCUSSION}
\smallskip

We have investigated the AC conductivity of binary random networks
of complex impedances,
with emphasis on its analytic structure in the complex variable $h$ or $\l$,
and on the corresponding resonant behaviour.
The present analysis is an extension to the general binary case
of a previous work~[2], devoted to the resonant response of a finite cluster,
or set of clusters, embedded in an infinite homogeneous host lattice.
Along the lines of refs.~[1, 2],
the poles of the conductance are interpreted in terms of the resonances
which show up in the AC conductivity of the $RL-C$ model,
and of the relaxation times in the transient response of the $R-C$ model.
We have proposed an efficient algorithm,
which allows a determination of the rational $\l$-dependence
of the conductance $Y$ of a finite sample,
in terms of its poles $\l_a$ and of the associated residues $\alpha_a$.
A very similar formalism had been proposed long ago by Straley~[6].

We want to underline again that the main advantage of the present approach
is to give at once the analytic structure
of the conductance in $h$ or $\l$, for any finite sample.
As far as a numerical investigation of the resonant response is concerned,
this approach is therefore more suitable than the usual numerical methods,
which can only yield the conductivity of the binary model for a fixed value
of the ratio $h$, such as the transfer-matrix method~[12],
or the iterative algorithm based on the $Y-\Delta$ transformation~[32],
used e.g. in ref.~[7].
It is also worth noticing
that our approach yields the full analytic structure of the conductance,
including the static conductance of the $\Q$-phase of any finite sample.
Indeed the latter quantity is given by eq.~(2.3),
with the corresponding residue $\alpha_1$ being given by the sum rule~(2.5)
in terms of all the other residues, corresponding to genuine resonances.

An extensive use of this algorithm, in the case of the square lattice,
has allowed us to investigate in detail
many aspects of the resonant dielectric response of the binary model.
In general we have used $10^7$ random bonds or more per measurement,
a good enough statistics to obtain very accurate data.
We have investigated the density of resonances $\rho(p,x)$
and the spectral function $H(p,x)$.
This approach yields a better evaluation of the spectral function than
the more direct method, consisting in measuring the imaginary part
of the conductivity.
The most salient structures have been labeled,
at least for a small enough concentration $p$,
by resonances of configurations made of one to three bonds,
which can be determined exactly, along the lines of ref.~[2].
The data for the spectral function have also been compared
with the EMA prediction.

The Lifshitz behaviour of the density of resonances near the endpoints
has been successfully characterised.
A good qualitative agreement is found with the argument of Hesselbo~[21],
according to whom the analogues of the Lifshitz sphere
are linear extended objects, such as hairpins.
From a quantitative viewpoint,
our data suggest that other classes of extended structures
may also contribute to the Lifshitz tails, even for a small concentration $p$.
The present situation is a lucky one,
since numerical investigations of Lifshitz tails
in more than one dimension are known to be a very difficult task in general,
especially in the case of binary disorder~[20].
The distribution of the spacings between neighbouring resonances
has been found in qualitative agreement with the universal distribution
of the GOE universality class of random matrices,
although small but definite differences show up at a quantitative level.
The number of visible resonances of the $RL-C$ model on a finite sample
of size $N\times N$, as a function of the quality factor $Q$,
has been shown to obey a finite-size scaling law~(4.5),
involving a scaling function $F(X)$ of the variable $X=Q/N^2$.
Quantities such as the scaling function $F(X)$,
or the spacing distribution $P(s)$, are expected not to be universal:
quantitative features of these functions
should rather weakly depend on the concentration $p$
and on the range of resonances considered,
the range~(4.3) being meant as a generic example.

More generally, for the binary model on a $d$-dimensional lattice,
the appropriate finite-size scaling variable describing dielectric resonances
reads $X=Q/N^d$.
Indeed, the number of resonances on a sample of linear size $N$ scales
as its volume $N^d$.
This observation yields in particular a prediction for
the divergence of the current correlation length $\xi$
in the weak-dissipation regime $(Q\gg 1)$.
By setting $X\sim 1$ for $N\sim\xi$, we get
$$
\xi\sim Q^{\overline{\nu}},\qquad\hbox{with}\quad\overline{\nu}=\frac{1}{d}.
\eqno(5.1)
$$
We thus recover a simple result due to Hesselbo~[21],
which has been corroborated by numerical simulations,
yielding $\overline{\nu}=0.4\pm0.1$ in two dimensions~[8].

Finally, we have investigated the distribution of the resonant electric field
living on any given bond.
The moments of this distribution obey power laws
with non-trivial exponents $x_k$, a characteristic feature of multifractality.
The associated generalised dimensions $d_k$ are found
to behave similarly to those observed in the Anderson model of localisation,
in the marginal two-dimensional case~[29].
Multifractality thus appears to be a quite generic feature of the resonant
response of binary networks.
In particular, this phenomenon is unrelated to the percolation transition.
In analogy with the fraction of visible resonances
or the spacing distribution $P(s)$,
the exponents $x_k$ and the dimensions $d_k$ are expected not to be
fully universal, but to exhibit a weak dependence on the concentration $p$,
and possibly on the range of resonances considered.

This multifractal picture provides a quantitative characterisation
of local features of the fluctuations in electric fields at resonance
observed previously~[8].
These giant fluctuations have been argued to be
responsible for surface-enhanced Raman scattering,
this phenomenon being especially pronounced
in strongly disordered semicontinuous films.
There is a regime where the enhancement factor is predicted
to be proportional to $\langle E^4\rangle$~[9],
whence the relevance of the dimension $d_4$, in our notation.
In a realistic system, with a small but finite dissipation rate $1/Q$,
these fluctuations are expected to be critical,
i.e., to exhibit strong spatial correlations,
on scales smaller than the current correlation length $\xi$,
estimated in eq.~(5.1).
Since the algorithm used in this work
gives direct access to the full map of electric fields at resonance,
it could be used to investigate other aspects of dielectric resonances,
including their spatial correlations,
on which some information is already available~[8].

\medskip
\noindent {\bf Acknowledgements}
\smallskip

It is a pleasure for us to thank
F. Brouers, J.P. Clerc, G. Giraud, and L. Raymond
for fruitful discussions,
and M.F. Thorpe for an interesting correspondence.

Part of this work has been done by Th. J. during his
{\it Stage de D.E.A. de Physique Quantique} (Paris).

\vfill\eject
{
\parindent 0pt
{\bf REFERENCES}
\bigskip

[1] J.P. Clerc, G. Giraud, J.M. Laugier, and J.M. Luck, Adv. Phys. {\bf 39}
(1990), 191.

[2] J.P. Clerc, G. Giraud, J.M. Luck, and Th. Robin, J. Phys. A {\bf 29}
(1996), 4781.

[3] F. Brouers, J.P. Clerc, and G. Giraud, Phys. Rev. B {\bf 44} (1991), 5299;
F. Brouers, J.P. Clerc, G. Giraud, J.M. Laugier, and Z.A. Randriamanantany,
Phys. Rev. B {\bf 47} (1993), 666;
F. Brouers, J.M. Jolet, G. Giraud, J.M. Laugier, and Z.A. Randriamanantany,
Physica A {\bf 207} (1994), 100, and references therein.

[4] D.J. Bergman, Phys. Rep. {\bf 43} (1978), 377;
Phys. Rev. Lett. {\bf 44} (1980), 1285;
Phys. Rev. B {\bf 23} (1981), 3058;
Ann. Phys. {\bf 138} (1982), 78.

[5] G.W. Milton, Appl. Phys. Lett. {\bf 37} (1980), 300;
J. Appl. Phys. {\bf 52} (1981), 5286; 5294;
Phys. Rev. Lett. {\bf 46} (1981), 542.

[6] J.P. Straley, J. Phys. C {\bf 12} (1979), 2143.

[7] A.R. Day and M.F. Thorpe, J. Phys. Cond. Matter {\bf 8} (1996), 4389.

[8] F. Brouers, S. Blacher, and A.K. Sarychev, in {\it Fractals in the
Natural and Applied Sciences}, N.M. Novak (ed.) (Chapman and Hall, 1995),
p. 237.

[9] F. Brouers, S. Blacher, A.N. Lagarkov, A.K. Sarychev, P. Gadenne, and V.M.
Shalaev, Phys. Rev. B {\bf 55} (1997), 13 234;
P. Gadenne, F. Brouers, V.M. Shalaev, and A.K. Sarychev, J. Opt. Soc. Am. B
{\bf 15} (1998), 68.

[10] H.A. Kramers and G.H. Wannier, Phys. Rev. {\bf 60} (1941), 252.

[11] J.P. Straley, Phys. Rev. B {\bf 15} (1977), 5733.

[12] J.M. Normand, H.J. Herrmann, and M. Hajjar, J. Stat. Phys. {\bf 52}
(1988), 441.

[13] J.M. Luck, Phys. Rev. B {\bf 43} (1991), 3933.

[14] D.A.G. Bruggeman, Ann. Phys. (Leipzig) [Folge 5] {\bf 24} (1935), 636.

[15] R. Landauer, J. Appl. Phys. {\bf 23} (1952), 779.

[16] S. Kirkpatrick, Rev. Mod. Phys. {\bf 45} (1973), 574.

[17] W.L. Moch\'an and R.G. Barrera (eds.),
Proceedings of {\it ETOPIM~3}, Physica A {\bf 207} (1994), 1--462;
A.M. Dykhne, A.N. Lagarkov, and A.K. Sarychev (eds.),
Proceedings of {\it ETOPIM~4}, Physica A {\bf 241} (1997), 1--452.

[18] I.M. Lifshitz, Adv. Phys. {\bf 13} (1964), 483;
Sov. Phys. -- Uspekhi {\bf 7} (1965), 549.

[19] Th.M. Nieuwenhuizen, Physica A {\bf 167} (1990), 43.

[20] M.C.W. van Rossum, Th.M. Nieuwenhuizen, E. Hofstetter, and M. Schreiber,
Phys. Rev. B {\bf 49} (1994), 13 377.

[21] B. Hesselbo, D. Phil. Thesis (Oxford, 1994).

[22] D. Stauffer and A. Aharony, {\it Introduction to Percolation Theory},
2nd ed. (Taylor and Francis, London, 1992).

[23] R.M. Ziff, S.R. Finch, and V.S. Adamchik, Phys. Rev. Lett. {\bf 79}
(1997), 3447.

[24] M.G. Watts, J. Phys. A {\bf 8} (1975), 61.

[25] O. Bohigas, in {\it Chaos and Quantum Physics}, Les Houches, Session LII
(M.J. Giannoni, A. Voros, and J. Zinn-Justin, eds.) (North-Holland, 1991).

[26] M.L. Mehta, {\it Random Matrices and Statistical Theory of Energy Levels}
(Academic Press, 1967; new revised and enlarged edition, 1990).

[27] G. Paladin and A. Vulpiani, Phys. Rep. {\bf 156} (1987), 147.

[28] J. Feder, {\it Fractals} (Plenum, New York, 1988).

[29] V.I. Fal'ko and K.B. Efetov, Europhys. Lett. {\bf 32} (1995), 627;
Phys. Rev. B {\bf 52} (1995), 17 413, and references therein.

[30] R. Rammal, C. Tannous, P. Breton, and A.-M.S. Tremblay, Phys. Rev. Lett.
{\bf 54} (1985), 1718;
R. Rammal, C. Tannous, and A.-M.S. Tremblay, Phys. Rev. A {\bf 31} (1985),
2662.

[31] L. de Arcangelis, S. Redner, and A. Coniglio, Phys. Rev. B {\bf 31}
(1985), 4725.

[32] C.J. Lobb and D.J. Franck, Phys. Rev. B {\bf 30} (1984), 4090;
D.J. Franck and C.J. Lobb, Phys. Rev. B {\bf 37} (1988), 302.

\vfill\eject
{\bf FIGURE CAPTIONS}
\bigskip

{\bf Figure 1:}
Schema of the binary network under consideration, with $M=N=8$.
The conductance $Y$ is measured between the plane electrodes (bus bars).
$\P$-bonds with conductance $\sigma_0$ are shown as solid lines,
$\Q$-bonds with conductance $\sigma_1$ are shown as dotted lines.

{\bf Figure 2:}
(a) hairpin configuration and (b) worm-like configuration, with $n=8$ cells.
Same conventions for the bond conductances as in Figure 1.
The transversal conductance $Y_n$ is measured between the point electrodes
shown as large dots.

{\bf Figure 3:}
Histogram plot of the spectral density of resonances $\rho(p,x)$,
for (a) $p=0.05$, (b) $p=0.1$, (c) $p=0.2$, (d) $p=0.3$, (e) $p=0.5$.
The dashed verticals show some of the resonances, listed in Table 1,
of the configurations made of one to three bonds, shown in Figure 5.

{\bf Figure 4:}
Histogram plot of $xH(p,x)$, with $H(p,x)$ being the spectral function
of the conductivity,
for (a) $p=0.05$, (b) $p=0.1$, (c) $p=0.2$, (d) $p=0.3$, (e) $p=0.5$.
Semi-circles show the EMA prediction~(2.53).
Dashed verticals as in Figure 3.

{\bf Figure 5:}
Configurations made of one to three bonds,
whose resonances are listed in Table 1,
and used to label the most salient peaks in Figures 3 and 4,
for $p$ small enough.

{\bf Figure 6:}
Logarithmic plot of the integrated density of resonances $\rho_\r{int}(p,x)$,
against $x^{-1/2}$, for a concentration $p=0.1$.
The slope of the least-squares fit yields the amplitude $C(p=0.1)\approx 3.19$.
Numbers label the lowest resonances of hairpin structures,
as explained in the text.

{\bf Figure 7:}
Plot of the measured amplitude $C(p)$ of the Lifshitz tail,
against $\abs{\ln p}$.
The fitted straight line is discussed in the text.

{\bf Figure 8:}
Histogram plot of the distribution $P(s)$ of normalised spacings
between neighbouring resonances,
measured in the range~(4.3).
The full line shows the GOE prediction.

{\bf Figure 9:}
Plot of the finite-size scaling function $F(X)$ of the fraction
of visible resonances, against the scaling variable $X=Q/N^2$.
Data correspond to samples of size $N\times N$,
with $N=12$ and $N=20$, at $p=1/2$.
The full line, showing the fit described in the text,
is meant as a guide for the eye.

{\bf Figure 10:}
Log-log plot of the moments $S_k(N)$ of the local electric field at resonance,
measured in the range~(4.3),
against the linear sample size $N$ $(6\le N\le 24)$.

{\bf Figure 11:}
Plot of the generalised dimensions $d_k$ of the distribution
of electric fields at resonance, measured in the range~(4.3),
against index $k$.
The straight line shows the linear behaviour~(4.12), with $\beta=0.194$.

{\bf Figure 12:}
Logarithmic histogram plot of the probability density $\Pi(\ell)$
of $\ell=\ln\abs{E}$, the logarithm of the electric field at resonance,
for samples of size $16\times 16$, measured in the range~(4.3).
The dashed vertical marks the upper bound $\ell_\max=(\ln n_B)/2$,
with $n_B=481$.

\vfill\eject
{\bf TABLE CAPTIONS}
\bigskip

{\bf Table 1:}
Location of the resonances of some configurations
consisting of one to three bonds, shown in Figure 5,
and used to label the most salient peaks in Figures 3 and 4,
for $p$ small enough.

{\bf Table 2:}
Exponents $x_k$ and associated generalised dimensions $d_k$
characterising the multifractal distribution of local electric fields
at resonance, measured in the range~(4.3).

\vfill\eject
}
\centerline{\bf Table 1}
\medskip
$$
\vbox{\init\halign to 13truecm
{\strut#&\vrule#\tabskip=1em plus 2em&
\hfil$#$\hfil&\vrule#&
\hfil$#$\hfil&\vrule#\tabskip 0pt\crr
&&\ &&\ &\cr
&&\hbox{configuration}&&\hbox{resonances}&\cr
&&\ &&\ &\crr
&&\ &&\ &\cr
&&\hbox{A}&&\l_{\r{A}1}=1/2=0.500000\hfill&\cr
&&\ &&\ &\crr
&&\ &&\ &\cr
&&\hbox{B}&&\matrix{
\l_{\r{B}1}=1/\pi=0.318310\hfill\cr
\l_{\r{B}2}=1-1/\pi=0.681690\hfill\cr}
\hfill&\cr
&&\ &&\ &\crr
&&\ &&\ &\cr
&&\hbox{C}&&\matrix{
\l_{\r{C}1}=1-2/\pi=0.363380\hfill\cr
\l_{\r{C}2}=2/\pi=0.636620\hfill\cr}
\hfill&\cr
&&\ &&\ &\cr
&&\hbox{D}&&\matrix{
\l_{\r{D}1}=\l_{\r{C}1}\hfill\cr
\l_{\r{D}2}=\l_{\r{C}2}\hfill\cr}
\hfill&\cr
&&\ &&\ &\crr
&&\ &&\ &\cr
&&\hbox{E}&&\matrix{
\l_{\r{E}1}=1/2-\sqrt{2}(1/2-1/\pi)=0.243051\hfill\cr
\l_{\r{E}2}=\l_{\r{A}1}\hfill\cr
\l_{\r{E}3}=1/2+\sqrt{2}(1/2-1/\pi)=0.756949\hfill\cr}
\hfill&\cr
&&\ &&\ &\crr
&&\ &&\ &\cr
&&\hbox{F}&&\matrix{
\l_{\r{F}1}=1/4+1/\pi-w/4=0.302436\hfill\cr
\l_{\r{F}2}=\l_{\r{C}1}\hfill\cr
\l_{\r{F}3}=1/4+1/\pi+w/4=0.834184\hfill\cr}
\hfill&\cr
&&\ &&\ &\cr
&&\hbox{G}&&\matrix{
\l_{\r{G}1}=3/4-1/\pi-w/4=0.165816\hfill\cr
\l_{\r{G}2}=\l_{\r{C}2}\hfill\cr
\l_{\r{G}3}=3/4-1/\pi+w/4=0.697564\hfill\cr}
\hfill&\cr
&&\ &&\ &\cr
&&\ &&\,w=\sqrt{9-40/\pi+48/\pi^2}\hfill&\cr
&&\ &&\ &\crr
}}
$$
\vfill\eject
\centerline{\bf Table 2}
\medskip
$$
\vbox{\init\halign to 10truecm
{\strut#&\vrule#\tabskip=1em plus 2em&
\hfil$#$\hfil&\vrule#&
\hfil$#$\hfil&\vrule#&
\hfil$#$\hfil&\vrule#\tabskip 0pt\crr
&&\ &&\ &&\ &\cr
&&k&&\hbox{exponent}{\hskip 5pt}x_k&&\hbox{dimension}{\hskip 5pt}d_k&\cr
&&\ &&\ &&\ &\crr
&&\ &&\ &&\ &\cr
&&0&&0&&2&\cr
&&\ &&\ &&\ &\crr
&&\ &&\ &&\ &\cr
&&1&&-0.097\pm0.016&&1.806\pm0.032&\cr
&&\ &&\ &&\ &\crr
&&\ &&\ &&\ &\cr
&&2&&0&&-&\cr
&&\ &&\ &&\ &\crr
&&\ &&\ &&\ &\cr
&&3&&\hfill0.295\pm0.018&&1.410\pm0.036&\cr
&&\ &&\ &&\ &\crr
&&\ &&\ &&\ &\cr
&&4&&\hfill0.769\pm0.026&&1.231\pm0.026&\cr
&&\ &&\ &&\ &\crr
&&\ &&\ &&\ &\cr
&&5&&\hfill1.390\pm0.026&&1.074\pm0.018&\cr
&&\ &&\ &&\ &\crr
&&\ &&\ &&\ &\cr
&&6&&\hfill2.130\pm0.024&&0.935\pm0.012&\cr
&&\ &&\ &&\ &\crr
}}
$$
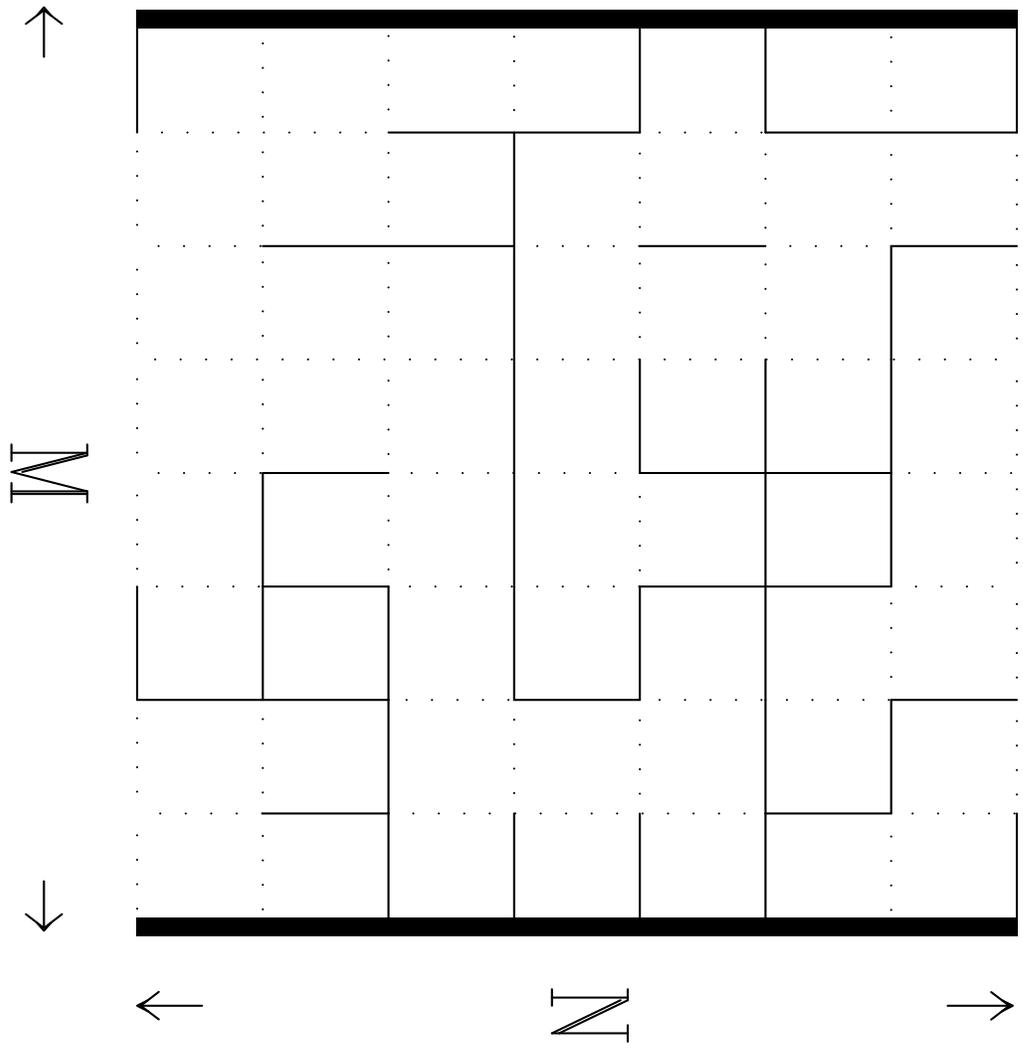
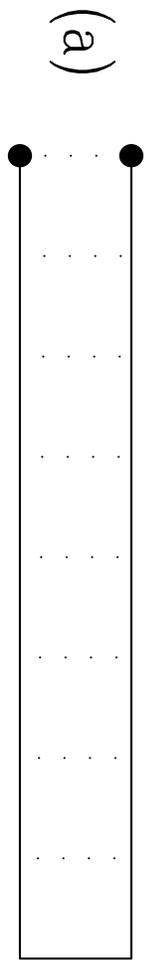
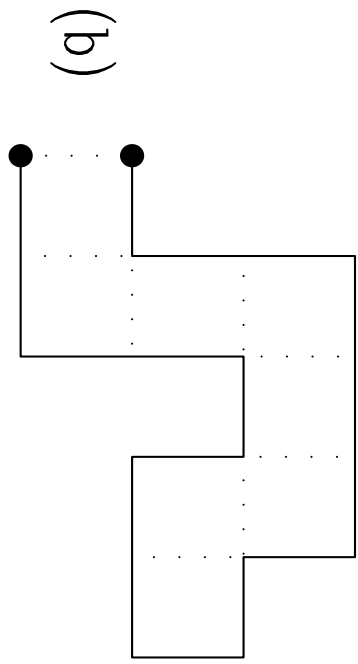
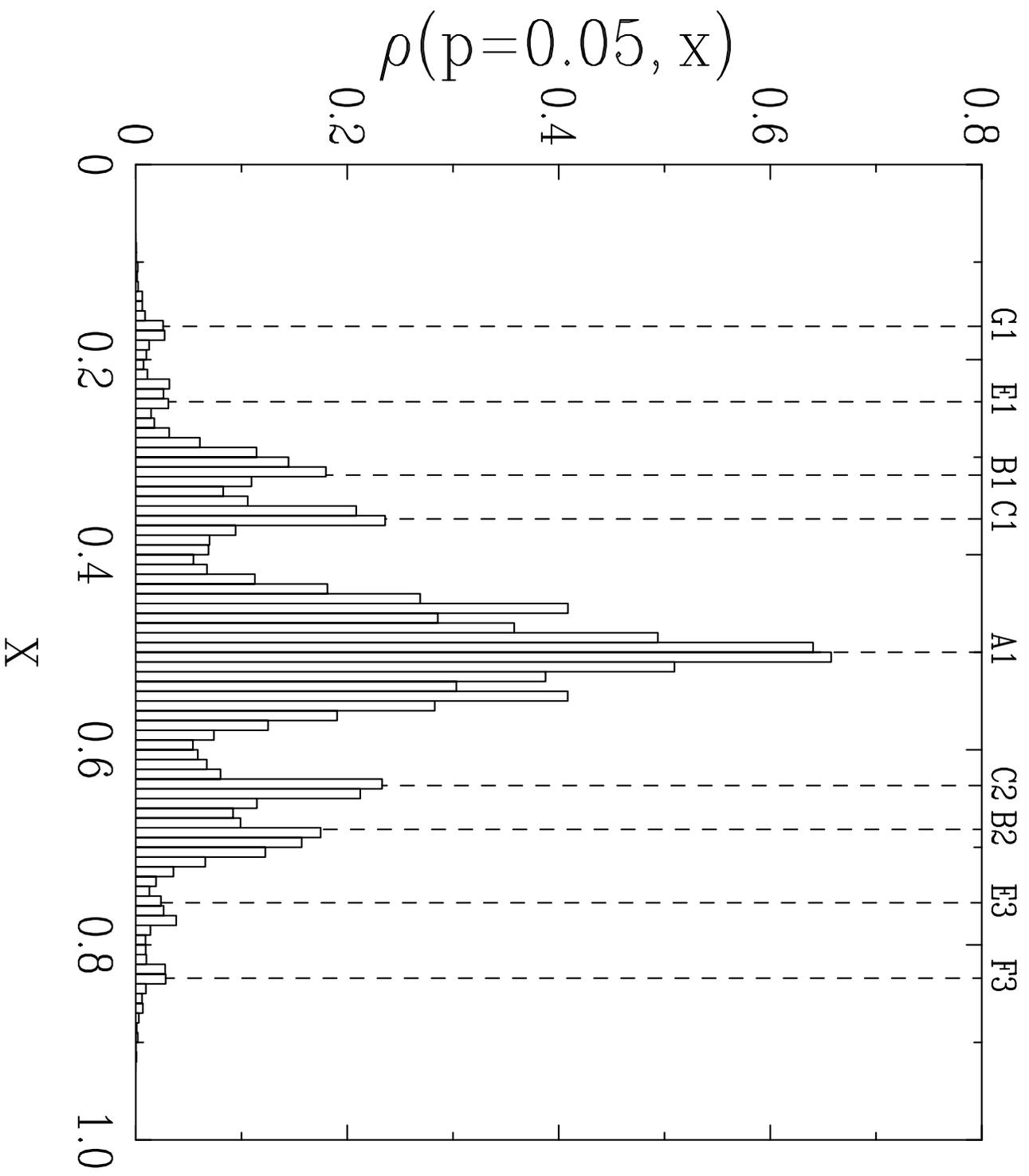
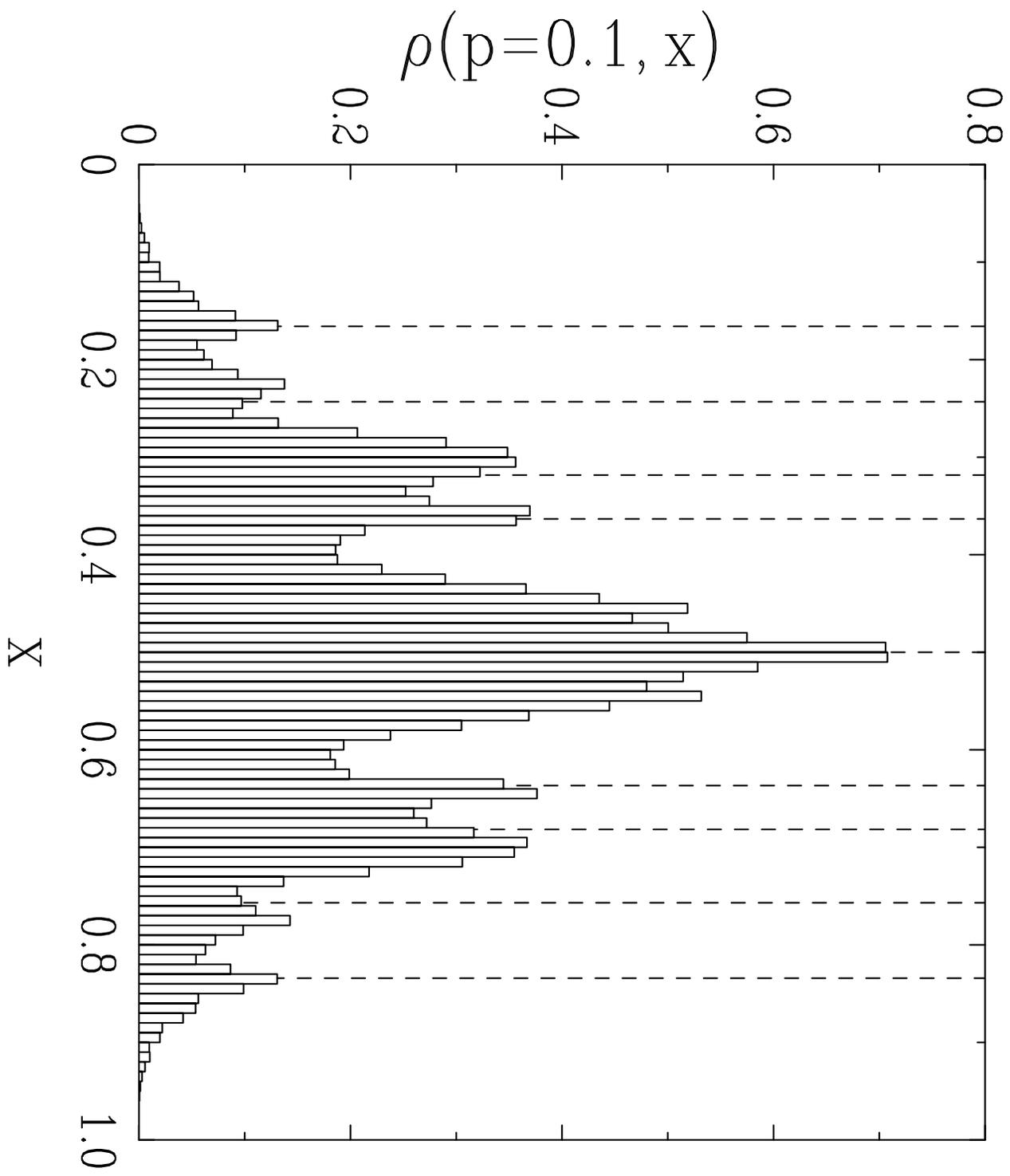
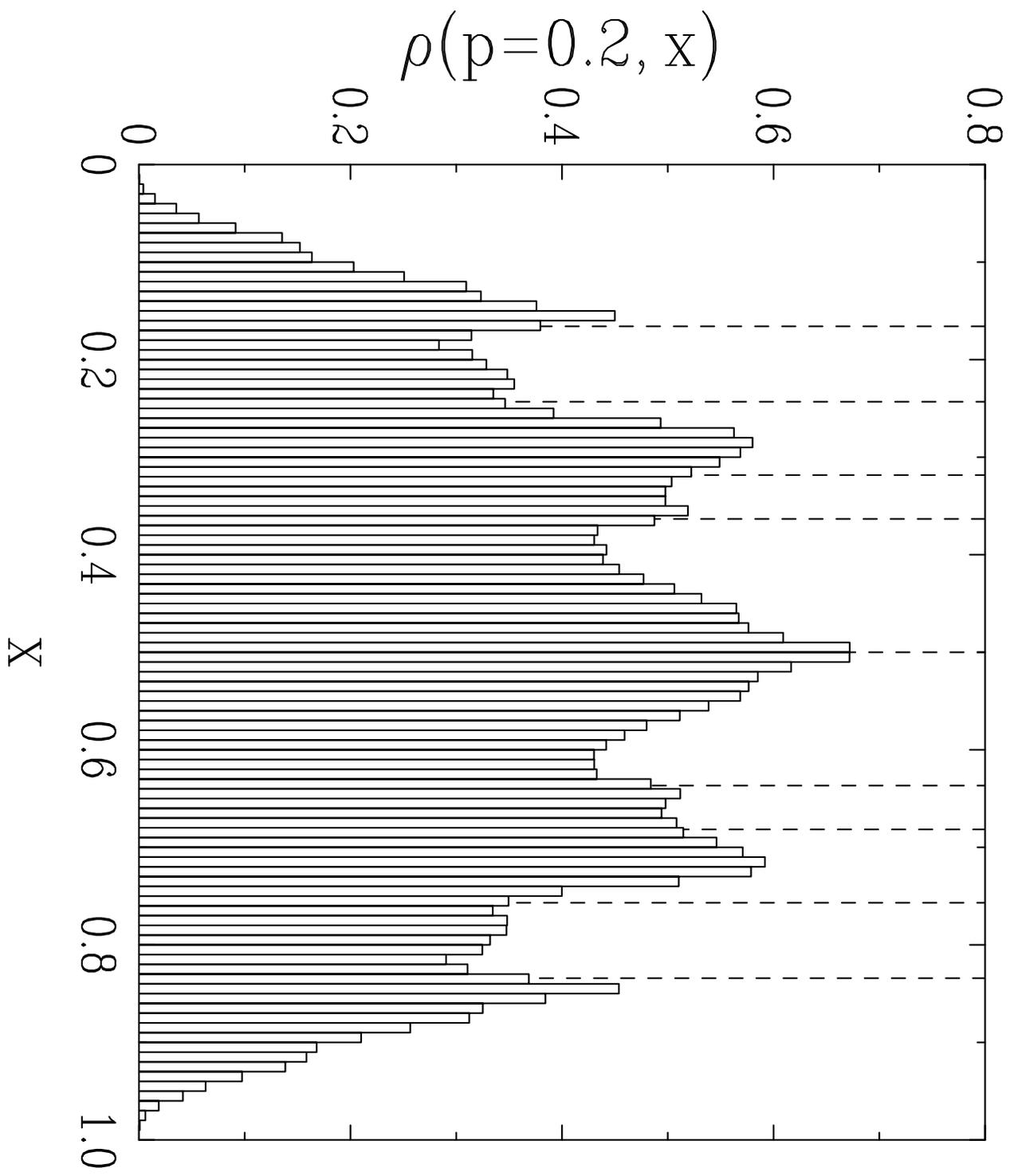
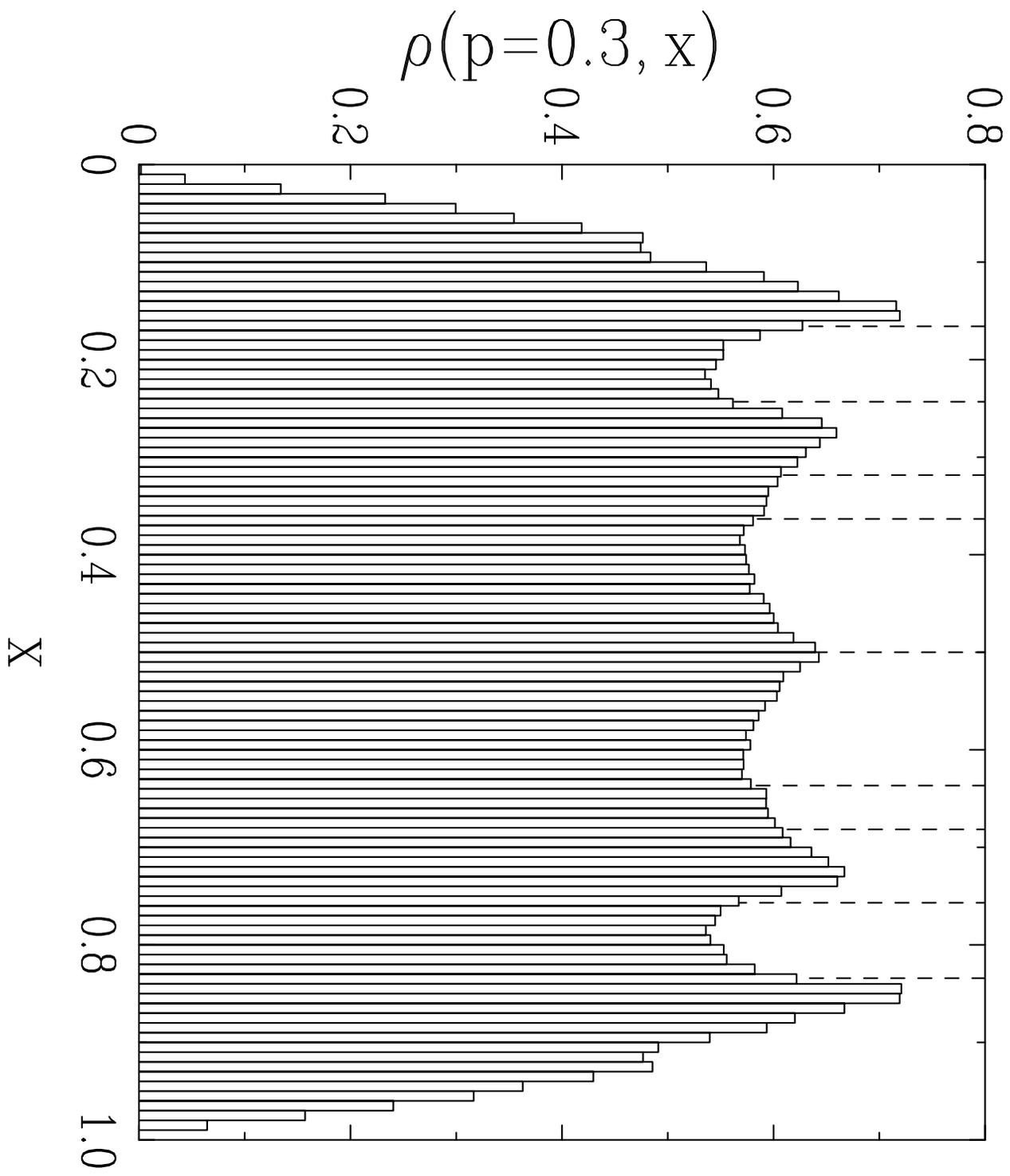
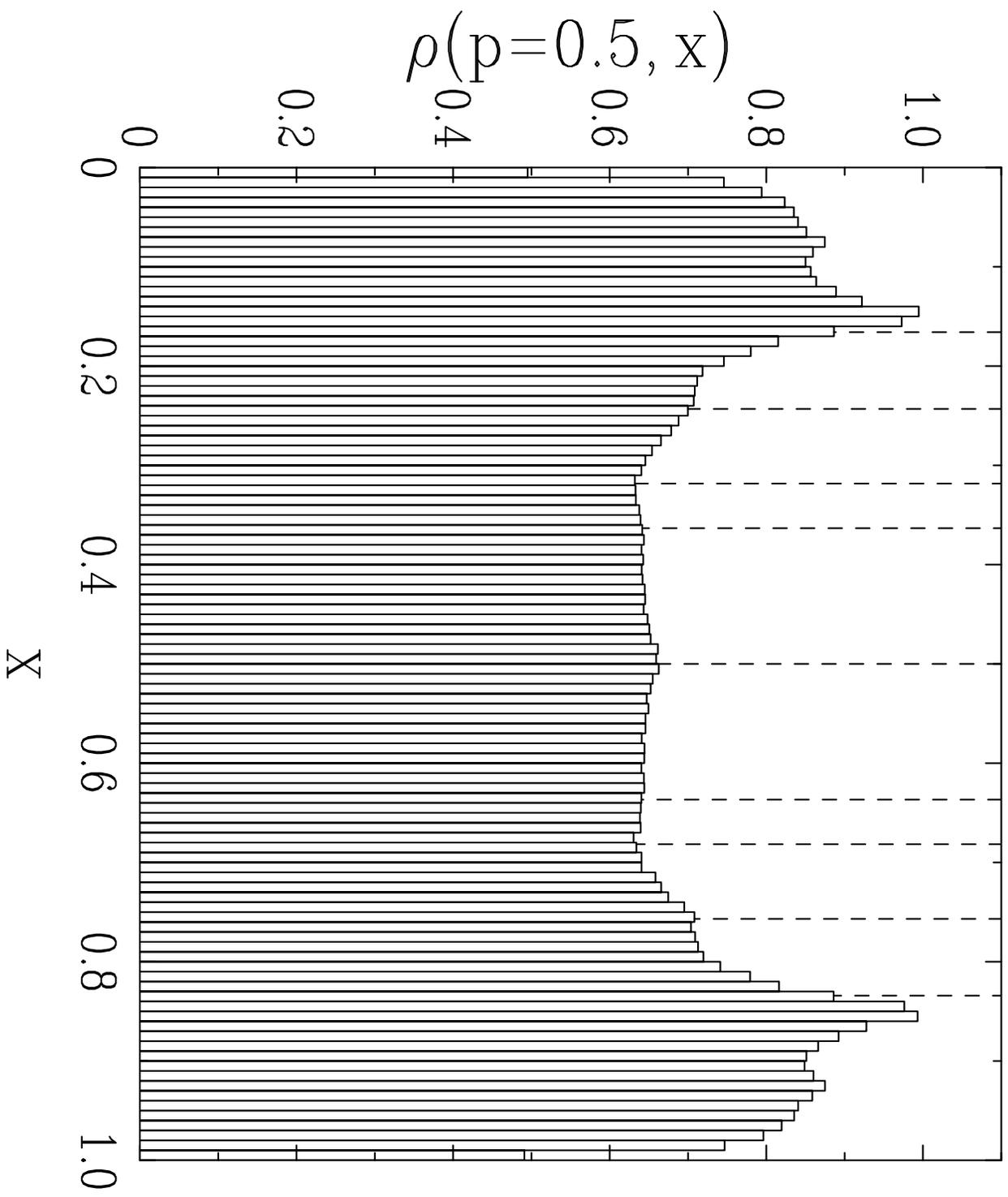
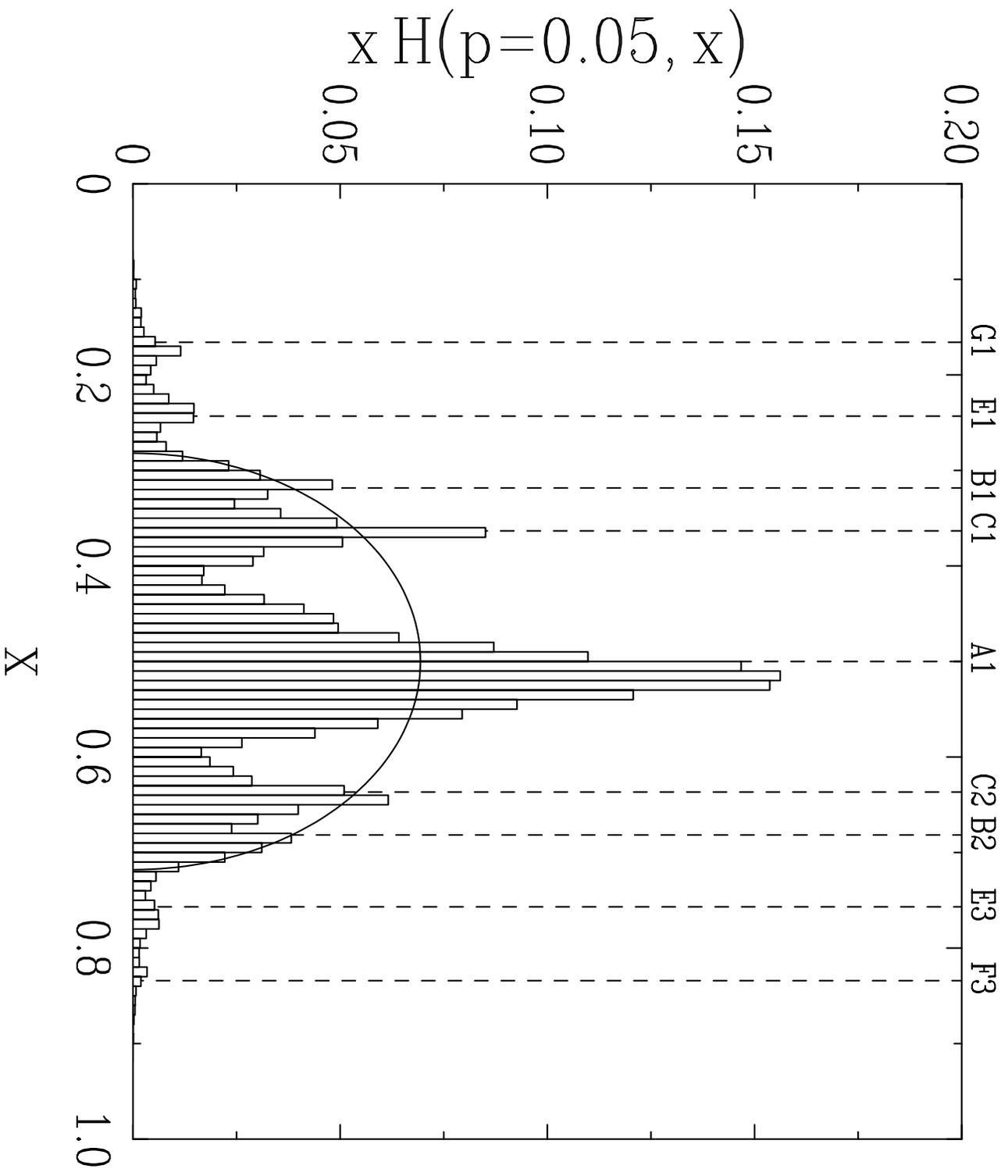
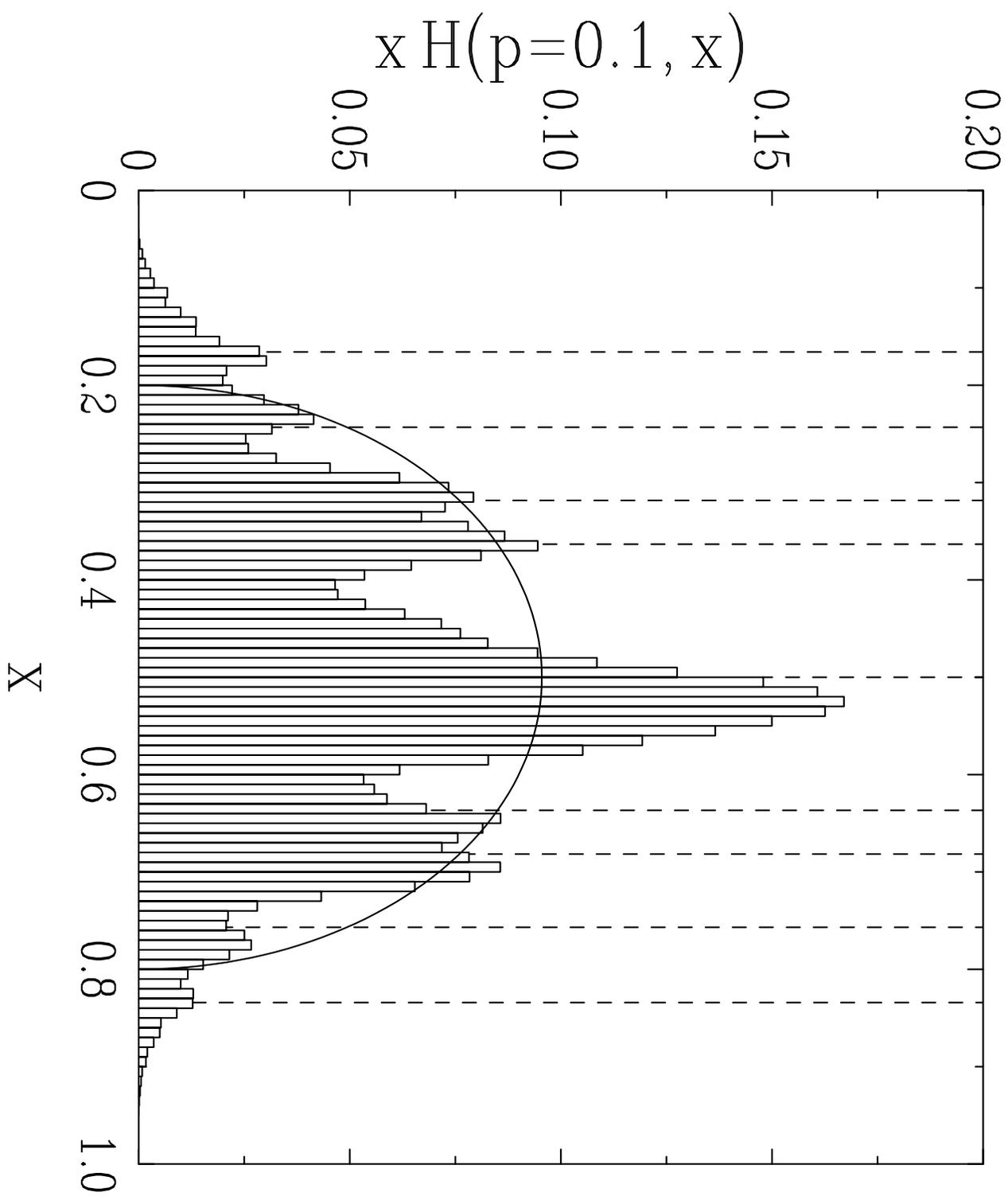
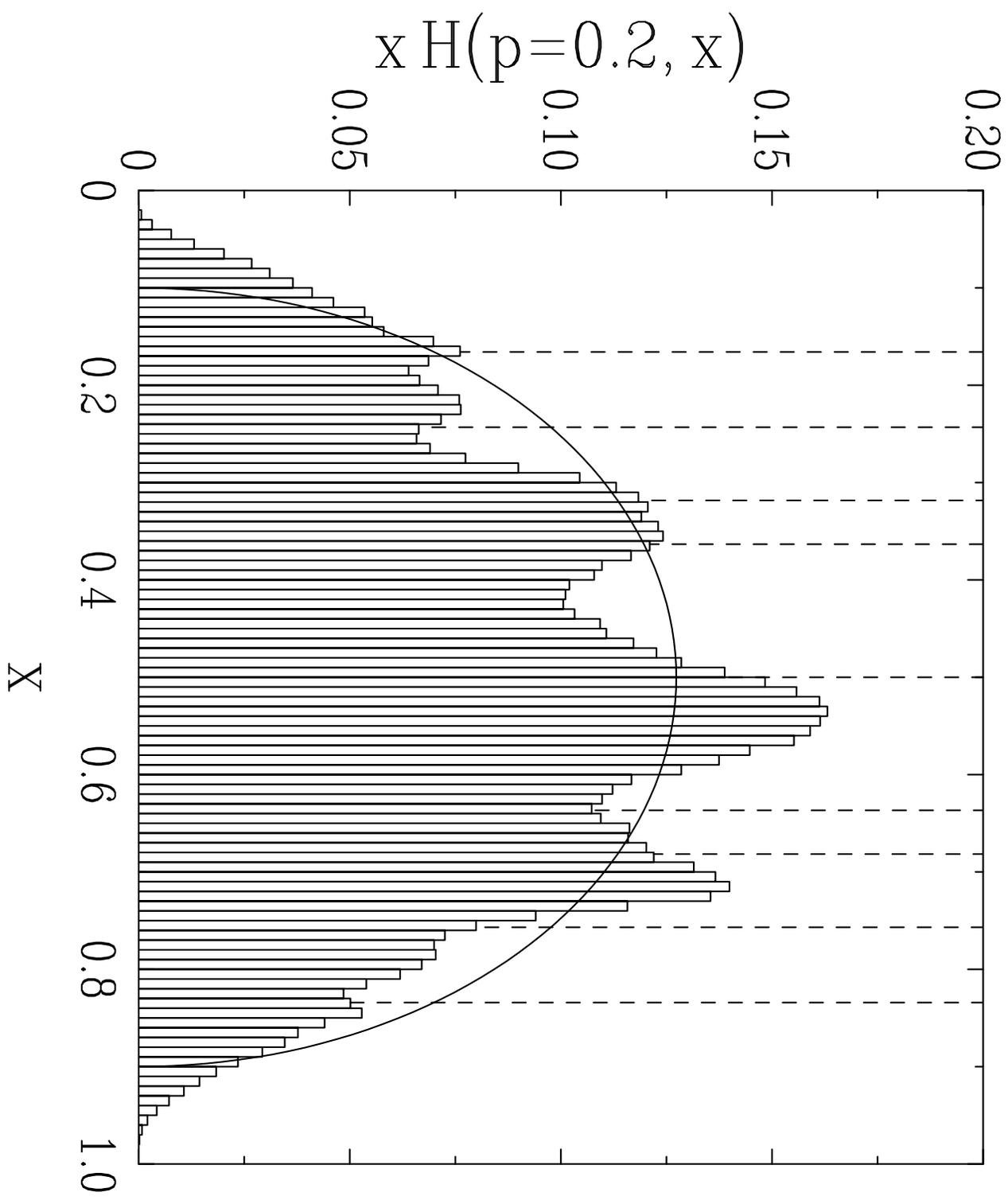
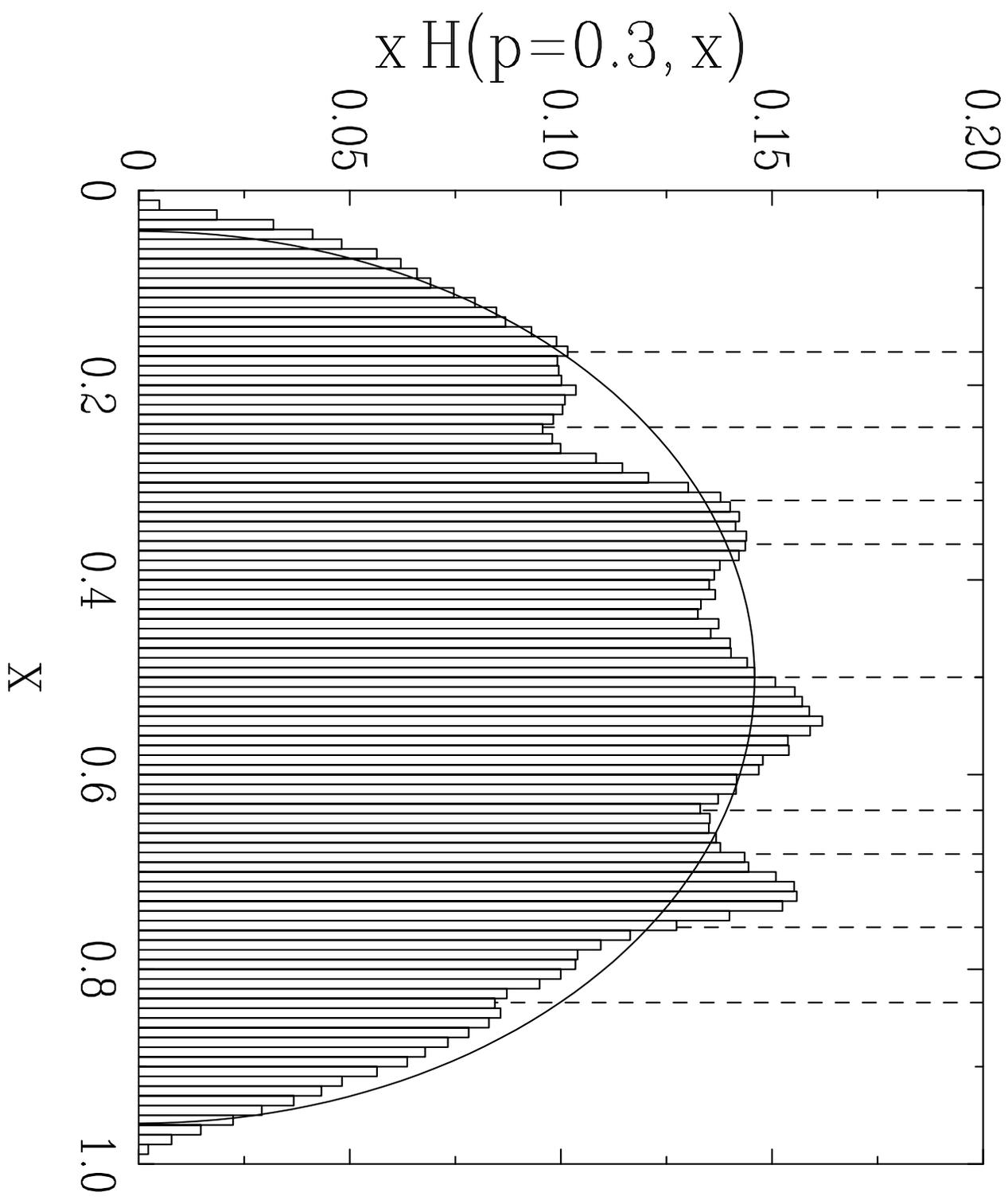
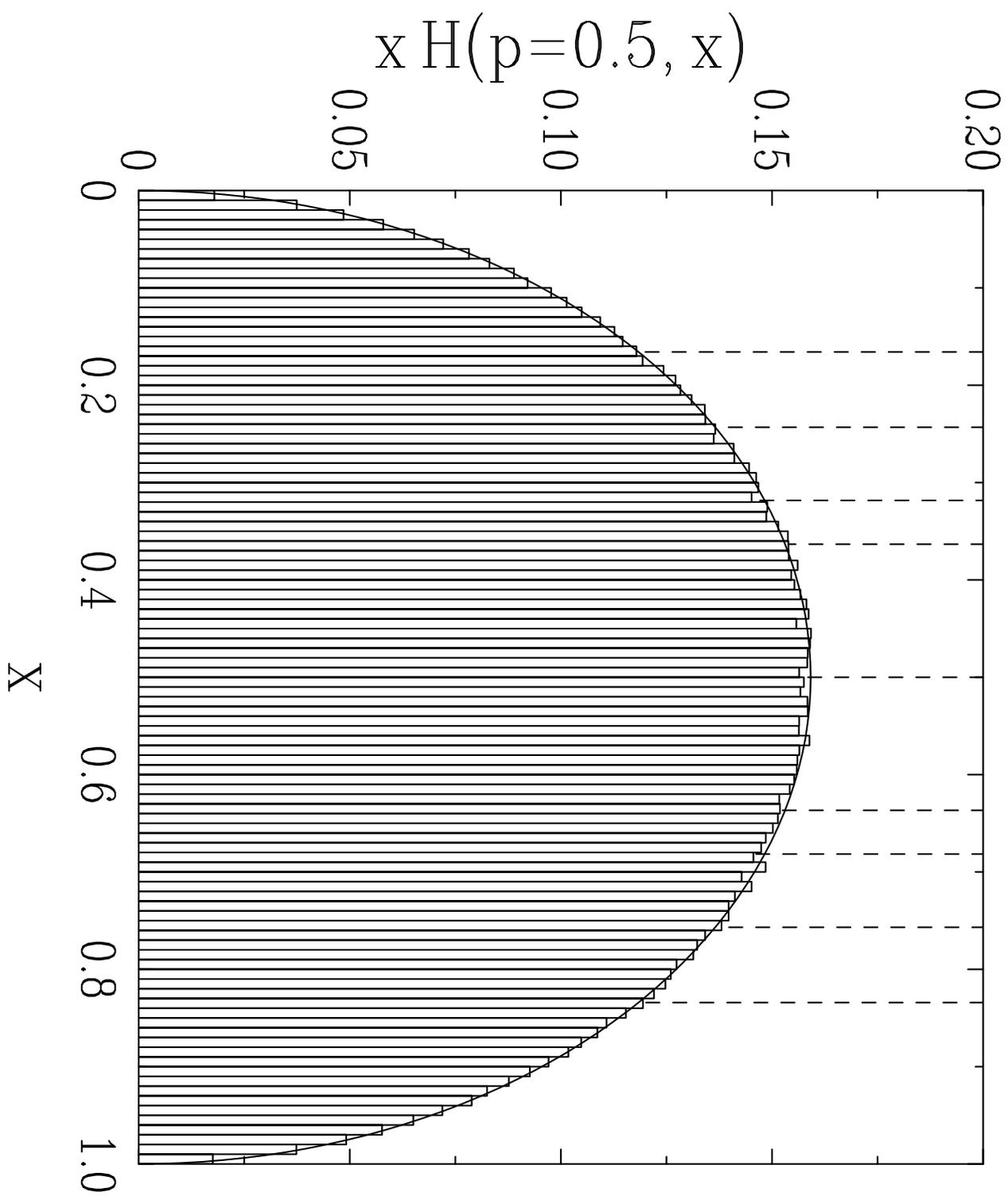
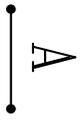
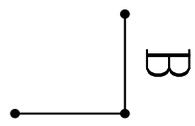
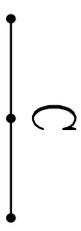
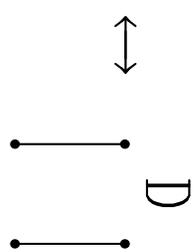
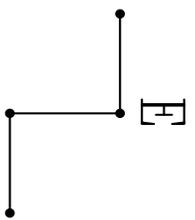
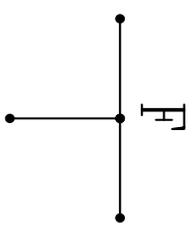
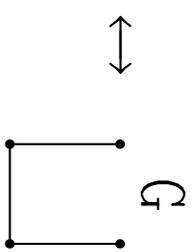
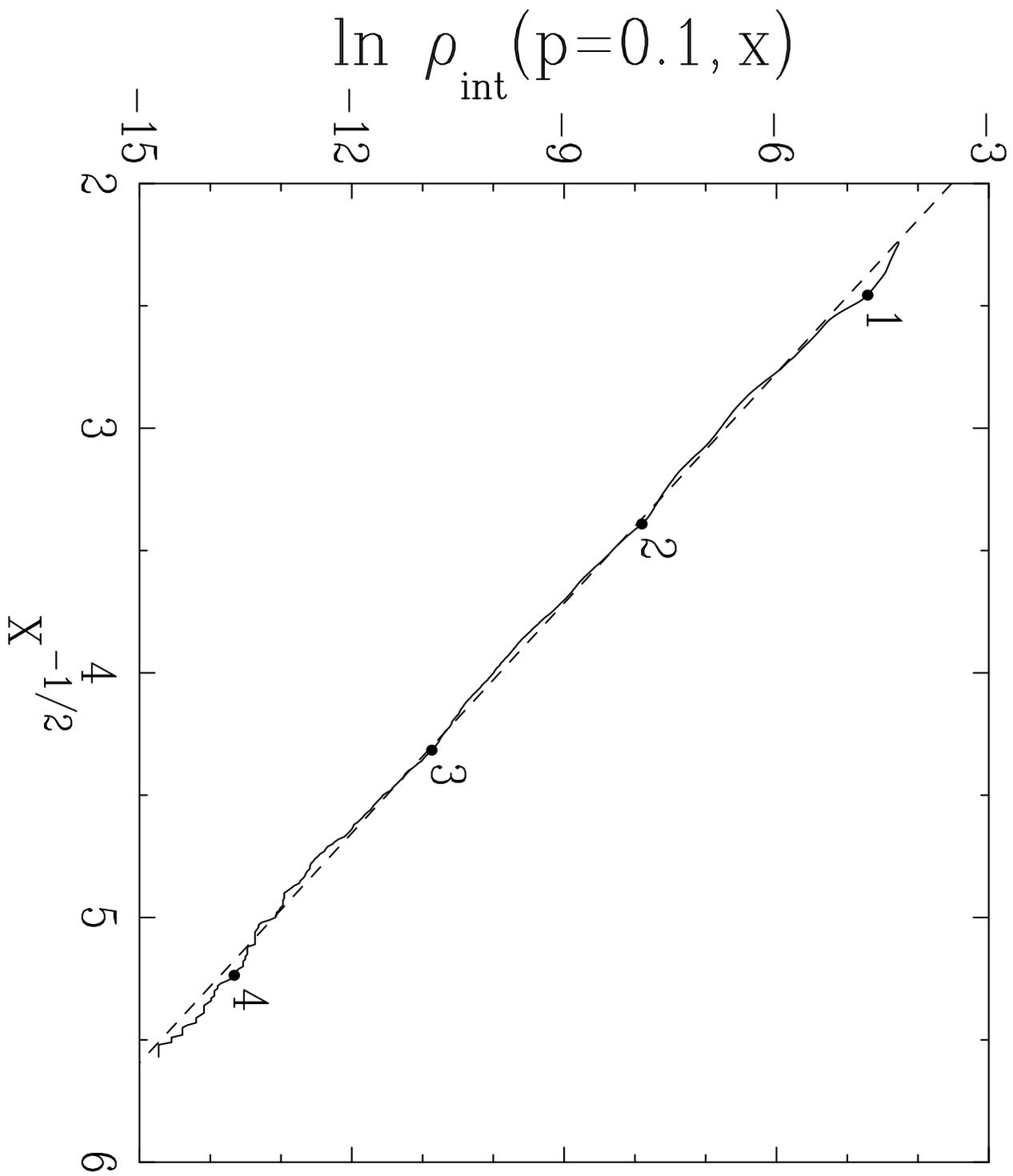
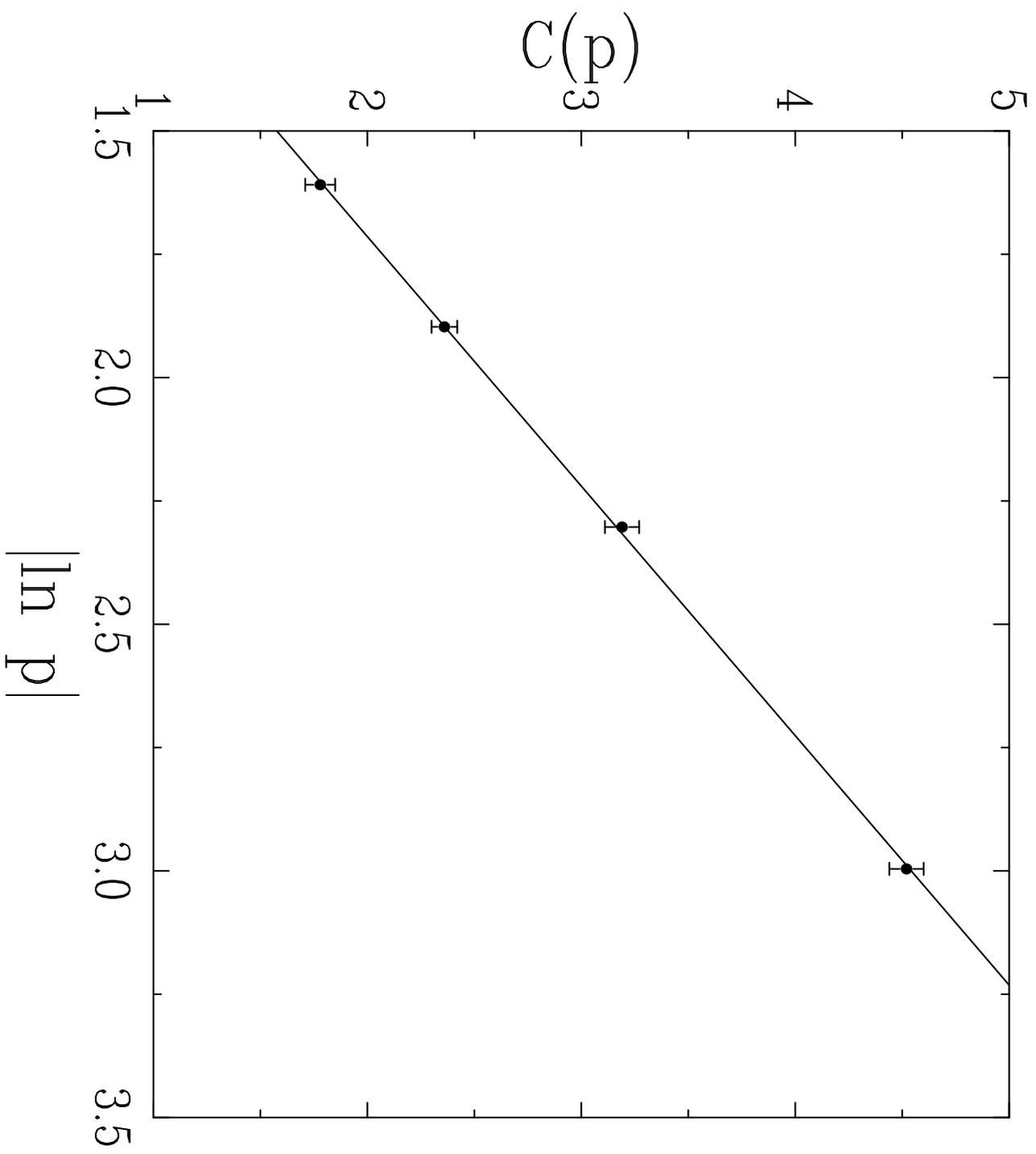
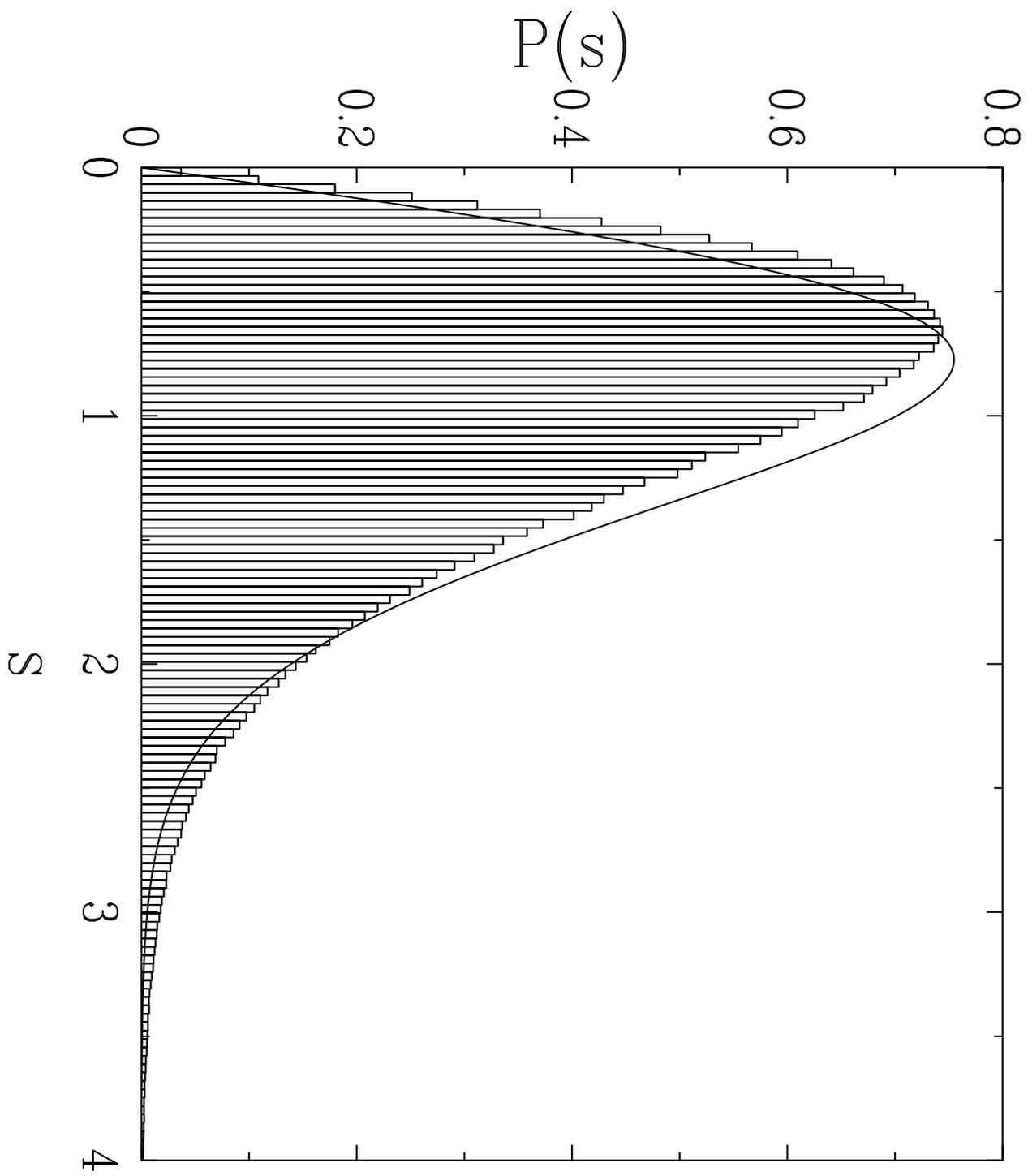
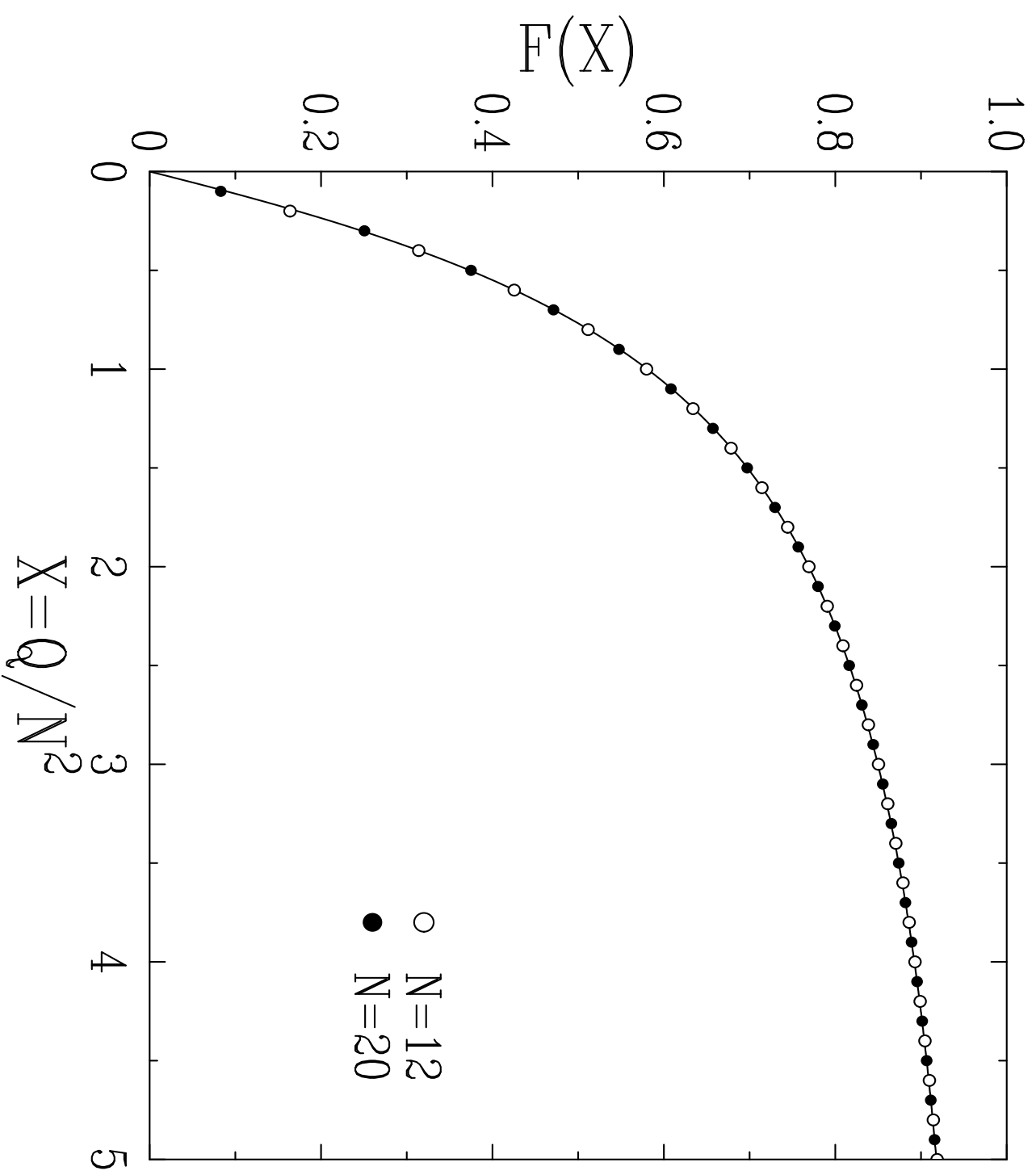
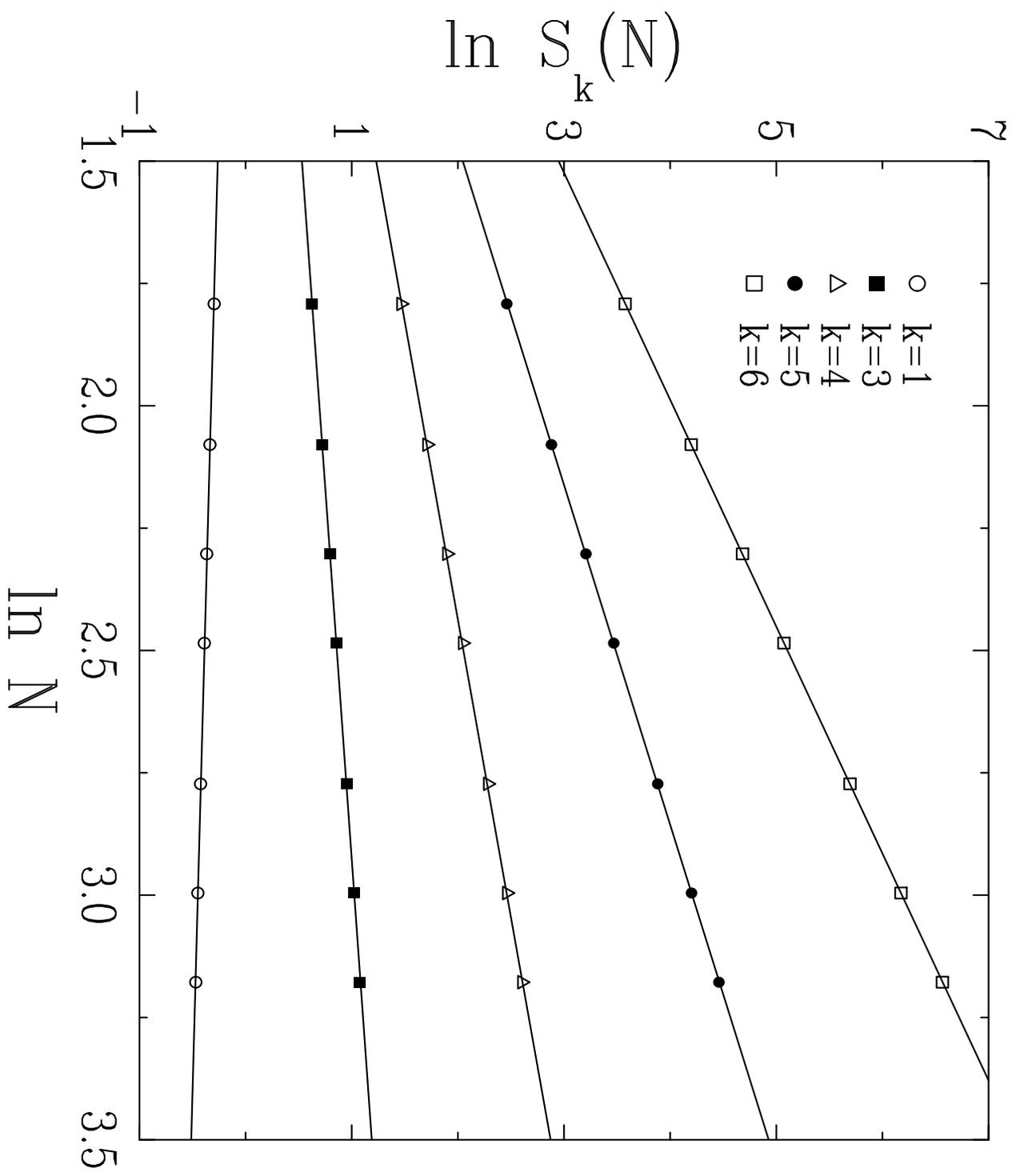
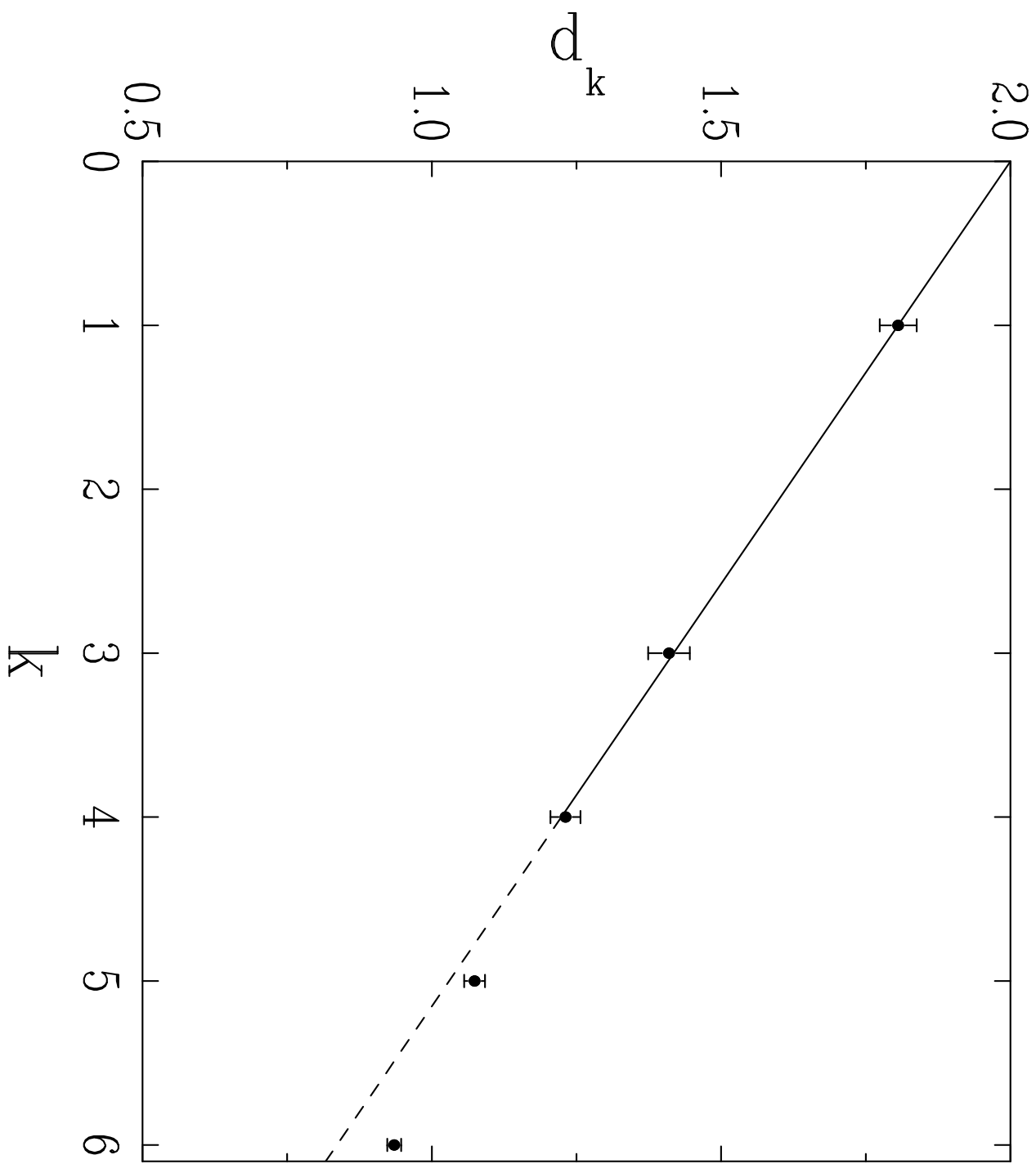
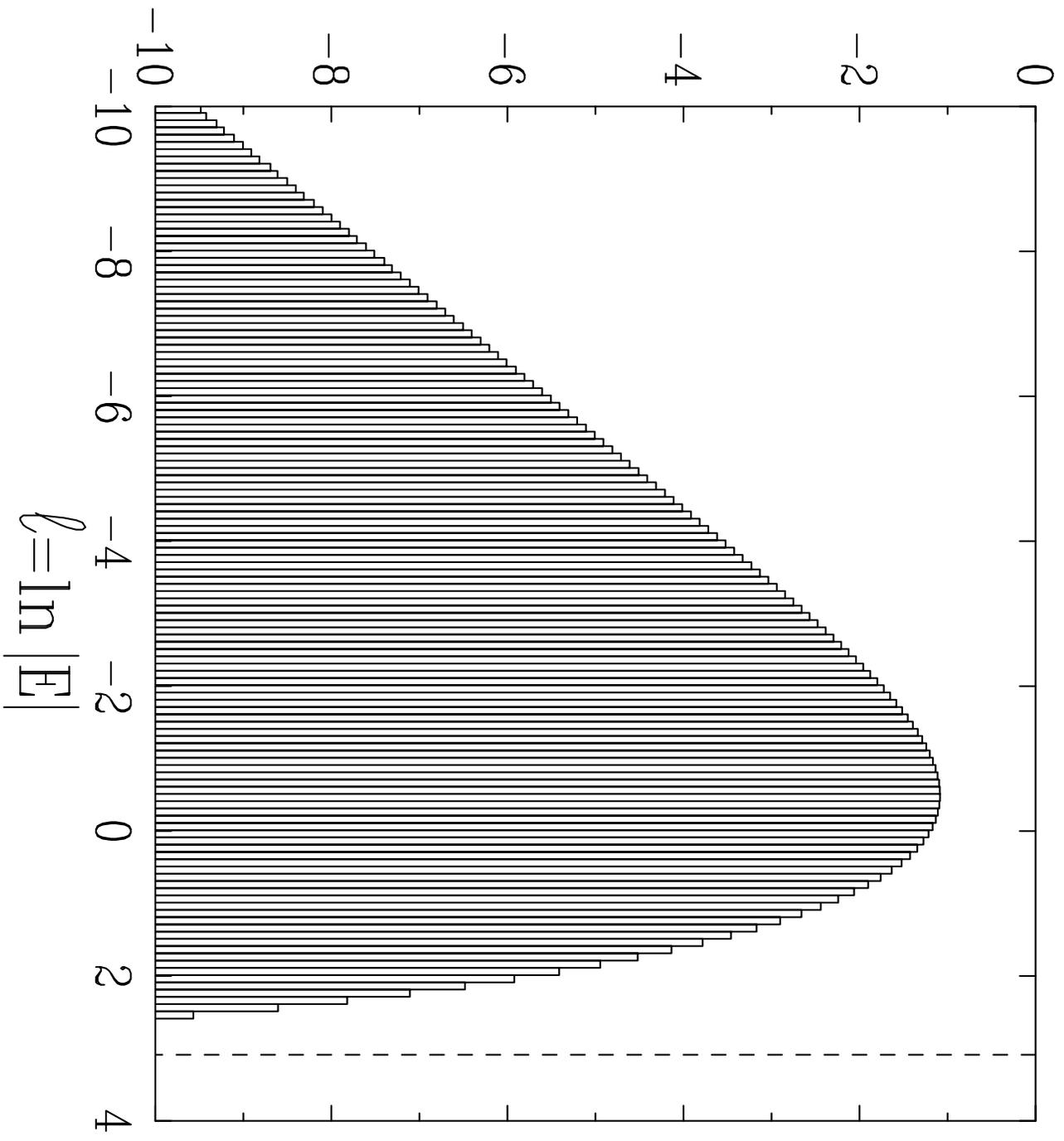
\bye